\documentclass[pdflatex,sn-mathphys-num]{sn-jnl}

\usepackage{tikz}
\usepackage{graphicx}
\usepackage{multirow}
\usepackage{lmodern}
\usepackage{bm} 
\usepackage{amsfonts}
\usepackage{amsthm}
\usepackage[title]{appendix}%
\usepackage{xcolor,colortbl}
\usepackage{textcomp}%
\usepackage{manyfoot}%
\usepackage{booktabs}%
\usepackage{algorithm}%
\usepackage{algorithmicx}%
\usepackage{algpseudocode}%
\usepackage{amsmath}
\usepackage{makecell}
\usepackage{multirow}
\usepackage{booktabs}
\usepackage[most]{tcolorbox}

\usepackage{subcaption}
\usepackage{diagbox}
\usepackage{tabularx}
\usepackage{siunitx}
\usepackage[T1]{fontenc}
\usepackage{hyperref}

\usepackage{booktabs} 
\newcolumntype{P}[1]{>{\raggedright\arraybackslash\hspace{0pt}}p{#1}}

\begin{document}

\title[Article Title]{How secure is AI-generated Code: A Large-Scale Comparison of Large Language Models}

\author[1,2]{\fnm{Norbert} \sur{Tihanyi}}

\author[3]{\fnm{Tamas} \sur{Bisztray}}

\author[4]{\fnm{Mohamed \sur{Amine Ferrag}}}

\author[2]{\fnm{Ridhi} \sur{Jain}}

\author[5,6]{\fnm{Lucas} \sur{C. Cordeiro}}

\affil[1]{ \orgname{Eötvös Loránd University (ELTE)}, \orgaddress{\city{Budapest}, \country{Hungary}}}

\affil[2]{ \orgname{Technology Innovation Institute (TII)}, \orgaddress{\city{Abu Dhabi}, \country{UAE}}}

\affil[3]{ \orgname{University of Oslo}, \city{Oslo}, \country{Norway}}

\affil[4]{\orgname{Guelma University}, \city{Guelma},  \country{Algeria}}

\affil[5]{\orgname{The University of Manchester}, \city{Manchester},  \country{UK}}

\affil[6]{\orgname{Federal University of Amazonas}, \city{Manaus},  \country{Brazil}}

\abstract{

This study compares state-of-the-art Large Language Models (LLMs) on their tendency to generate vulnerabilities when writing C programs using a neutral zero-shot prompt. Tihanyi et al. introduced the FormAI dataset at PROMISE '23, featuring 112,000 C programs generated by GPT-3.5-turbo, with over 51.24\% identified as vulnerable. We extended that research with a large-scale study involving 9 state-of-the-art models such as OpenAI's GPT-4o-mini, Google's Gemini Pro 1.0, TII's 180 billion-parameter Falcon, Meta's 13 billion-parameter Code Llama, and several other compact models. Additionally, we introduce the FormAI-v2 dataset, which comprises 331$\,$000 compilable C programs generated by these LLMs. Each program in the dataset is labeled based on the vulnerabilities detected in its source code through formal verification, using the Efficient SMT-based Context-Bounded Model Checker (ESBMC). This technique minimizes false positives by providing a counterexample for the specific vulnerability and reduces false negatives by thoroughly completing the verification process. Our study reveals that at least 62.07\% of the generated programs are vulnerable. 
The differences between the models are minor, as they all show similar coding errors with slight variations. Our research highlights that while LLMs offer promising capabilities for code generation, deploying their output in a production environment requires proper risk assessment and validation. \textbf{Please cite this once published: \url{https://doi.org/10.1007/s10664-024-10590-1}}.
}


\maketitle

\section{Introduction}\label{sec:introduction}

Large Language Models (LLMs) are transforming software development and programming~\cite{wang2024software, xu2022systematic, jain2022jigsaw}. Every day, developers and computer scientists utilize various code creation and completion models to tackle different tasks~\cite{bui_codetf_2023,ross_programmers_2023}. Research related to program synthesis using Generative Pre-trained Transformers (GPT)~\cite{chavez_chat_2023} is gaining significant traction, where initial studies indicate that GPT models can generate syntactically correct yet vulnerable code~\cite{charalambous_new_2023}.

A study conducted at Stanford University suggests that software engineers assisted by \textit{OpenAI’s codex-davinci-002 model} were at a higher risk of introducing security flaws into their code~\cite{perry_users_2022}. As the usage of AI-based tools in coding continues to expand, understanding their potential to introduce software vulnerabilities becomes increasingly important. Given that LLMs are trained on data freely available on the internet, including potentially vulnerable code, there is a high risk that AI tools could replicate the same patterns. This raises a critical question: \textit{Is it safe to employ these models in real projects?} 

As a first step towards answering this question, Tihanyi et al. published the FormAI dataset~\cite{FormAI} at the 19th International Conference on Predictive Models and Data Analytics in Software Engineering (PROMISE'23). This dataset is the first and largest collection of AI-generated compilable C programs with vulnerability classification, featuring $112\,000$ samples. To guarantee the diversity of generated C codes, the authors developed a framework designed to produce a variety of programs that cover multiple coding scenarios, efficiently facilitating real-world bugs. The study employed Bounded Model Checking (BMC), a technique within Formal Verification (FV), to evaluate the security properties of the dataset. This initial study revealed that at least $51.24$\% of the C programs generated by GPT-3.5-turbo were vulnerable. 

Continuing the original research presented in \cite{FormAI}, we aim to expand the scope of the study by addressing the key limitations highlighted by the research community. We identified four main limitations that we intend to address from the original paper:
\begin{enumerate}

\item The first paper exclusively focuses on OpenAI's GPT-3.5-turbo, without evaluating other models. To bridge this gap, this paper compares nine state-of-the-art LLMs in secure coding, such as Google's \textit{Gemini Pro 1.0}~\cite{team2023gemini}, OpenAI's GPT-4o-mini, TII's Falcon-180B~\cite{almazrouei2023falcon}, and Meta's Code LLama 13B~\cite{roziere2023codellama}. In addition, we have expanded the original dataset from $112\,000$ to $331\,000$ samples, where incorporating C code generated by different LLMs also enhances diversity.
\item The initial findings on the percentage of vulnerable samples in the dataset ($51.24$\%) may have been under-reported due to the limitations of bounded model checking, indicating that the actual percentage of vulnerabilities could be higher. To address this issue, we transitioned our verification approach from bounded to \textit{unbounded} verification, thereby enhancing the depth and accuracy of our security evaluation~\cite{GadelhaIC17,gadelha2018esbmc,GadelhaMCN19,MenezesAFLMSSBGTKC24}.
\item We have incorporated new labels into the dataset to enhance its usability for a broader research community. While all necessary features can be extracted and reproduced directly from the provided source codes, we have enhanced the dataset's comprehensiveness by calculating the \textit{cyclomatic complexity} (CC)~\cite{McCabe} for each program, adding source lines of code (SLOC), including the exact stack trace for counterexamples, and providing a code snippet entry that captures only the 5 lines before and after the vulnerability. These additional features are valuable for machine learning tasks to help models generalize the problem and identify vulnerabilities more effectively and conduct more detailed comparisons in various research contexts.
\item  To enhance the dataset, we have removed all Type 1, Type 2, Type 3-1, and Type 3-2 (with 10\% deviation threshold) clones using the NiCad (Automated Detection of Near-Miss Intentional Clones)~\cite{Cordy2011TheNC} tool. We note, that removing Type 3-2 clones with a larger threshold is not our goal. Even minor changes can be significant and determine whether a vulnerability is present or absent, potentially introducing different security risks. Moreover, different representations of a vulnerability can help models better generalize during machine learning training.
\end{enumerate}
This study answers the following research questions:

\begin{tcolorbox}[colback=gray!10]

\begin{itemize}
\item {\textbf{RQ1}:} How does the security of LLM-generated code differ across various models? 
\item {\textbf{RQ2}:} What are the most typical vulnerabilities introduced during C code generation by different LLMs using neutral zero-shot prompts?
    
\end{itemize}

\end{tcolorbox}

\subsection{Main contribution}
To summarize, this paper holds the following original contributions:
\begin{itemize}

    \item We present the \texttt{FormAI-v2} dataset, consisting of $331\,000$ compilable C programs ($310 \,531$ with the exclusion of any Type 1, Type 2, Type 3-1 and Type 3-2 clones) generated by nine different LLMs. Each C sample has been systematically labeled based on vulnerabilities identified through formal verification methods, particularly using the Efficient SMT-based Bounded Model Checker (ESBMC)~\cite{gadelha2018esbmc,GadelhaMCN19,MenezesAFLMSSBGTKC24} tool with an \textit{unbounded} setting;
    
    \item A detailed study to determine which models produce code with the highest and the lowest number of vulnerabilities;

    \item We provide a comprehensive analysis of the generated programs, detailing the distribution of vulnerabilities and highlighting the most frequently encountered types;
    \item We made the \texttt{FormAI-v2} dataset available to the research community, including all generated C samples and classification results. The dataset can be accessed on our project website at \url{https://github.com/FormAI-Dataset}.

\end{itemize}

The remaining sections are organized as follows: Section~\ref{sec:motivation} provides an in-depth discussion on the motivation. Section~\ref{sec:related} presents a comprehensive overview of the literature related, highlighting significant previous studies and their findings. Section~\ref{sec:formal} introduces the concepts of formal verification, focusing on the ESBMC module. Section~\ref{sec:methodology} details the methodology we adopted to develop and label our dataset. Section~\ref{sec:Discussion} presents our findings and discusses their implications. Section~\ref{sec:futureandlimit} explores the limitations and threats to the validity, and proposes potential future research directions. Finally, Section~\ref{sec:conclusion} concludes the paper by summarising our contributions and addressing the research questions posed in this study.

\section{Motivation}
\label{sec:motivation}

\begin{figure}[ht]
    \includegraphics[width=1\textwidth]{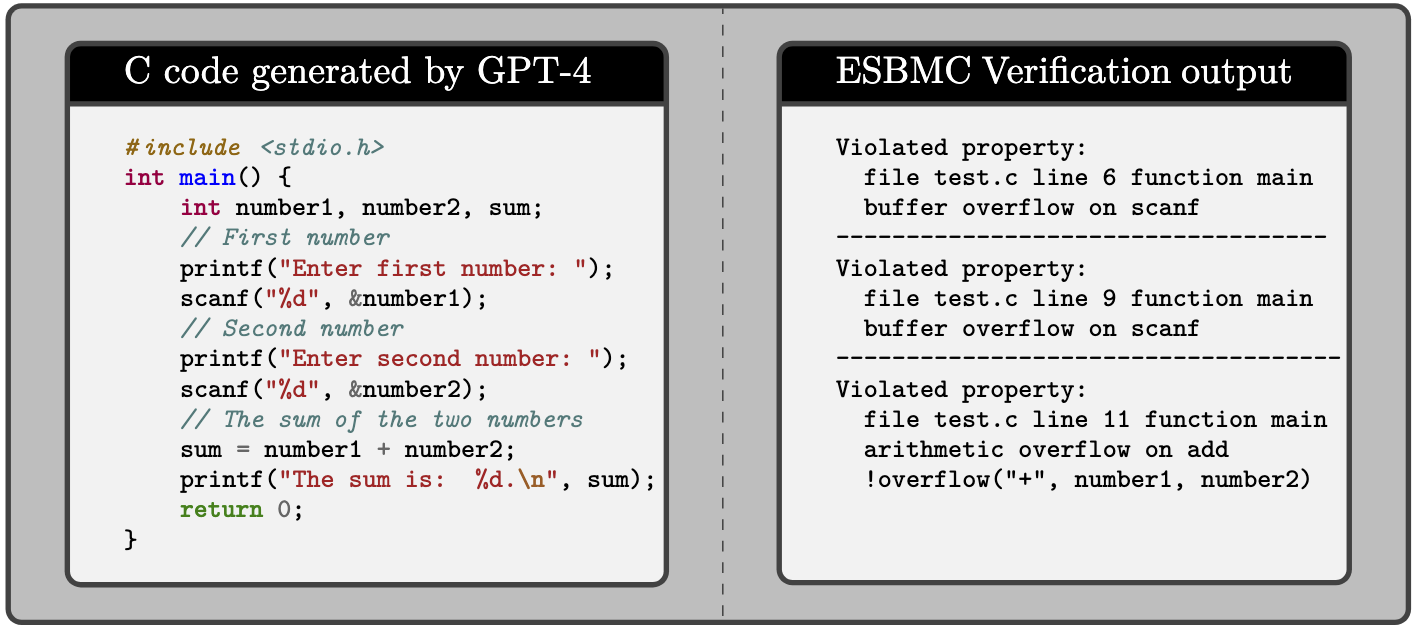}
    \caption{Motivation example: GPT-4 produced code with security vulnerabilities,
demonstrated through formal verification results.}
\label{fig:gptmotivation}

\end{figure}

In program synthesis, LLMs are generally used for simple tasks like writing a prime number generator or a basic program to sort an array, rather than handling large-scale projects involving thousands of lines of code~\cite{perry_users_2022}. The latest generation of LLMs can easily solve these simple tasks without facing any challenges. So far, the main area of interest in LLM-based code generation has been correctness. Datasets such as HumanEval~\cite{chen2021codex} provide programming challenges to assess the performance of models in correctly solving various problems. For example, GPT-4 achieves a $67\%$ success rate in solving tasks compared to $48.1\%$ for GPT-3.5-turbo~\cite{openai_gpt-4_2023}.
On the contrary, even for basic programming tasks, state-of-the-art LLMs may adopt insecure coding practices. To illustrate the issue, imagine a situation where a programmer asks GPT-4 the following: \textit{``Create a C program that prompts the user to input two numbers and then calculate their sum''}. The code generated by GPT-4 is presented on the left in Figure~\ref{fig:gptmotivation}, while the output from the formal verification tool ESBMC 7.6.1 is shown on the right.

\noindent This simple code contains three potential security vulnerabilities. It exhibits an integer overflow during the addition of the variables \textit{number1} and \textit{number2}, as well as two buffer overflows through the \texttt{scanf()} functions that retrieve input from the user.
In $32$-bit computing architectures, integers are commonly stored as $4$ bytes ($32$ bits), which results in a maximum integer value of $2\,147\,483\,647$, equivalent to $2^{31}-1$. If one attempts to add $2\,147\,483\,647 + 1$ using this small program, the result will be incorrect due to integer overflow.
The incorrect result will be $-2\,147\,483\,648$ instead of the expected $2\,147\,483\,648$. The addition exceeds the maximum representable value for a signed 32-bit integer $2^{31}-1$, causing the integer to wrap around and become negative due to the two's complement representation.

When GPT-4 is requested to write a secure version of this code using the following prompt: \textit{``Create a C program that prompts the user to input two numbers and then calculates their sum. Be careful and avoid security vulnerabilities.''}, it only attempts to fix entering non-integer inputs by adding the following code snippet (Figure~\ref{fig:2}): 

\begin{figure}[ht]

\includegraphics[width=1\textwidth]{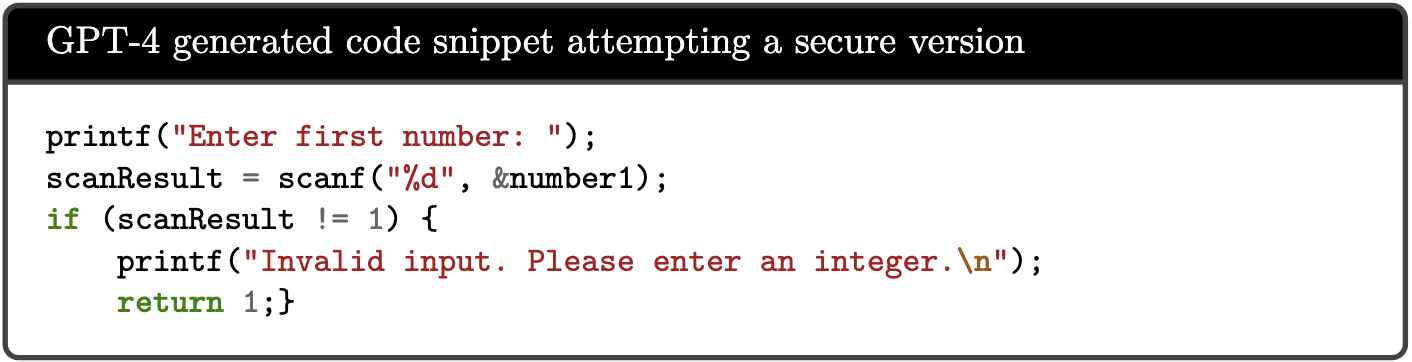}
\caption{GPT-4 generated code snippet response after requesting a secure version of the code in Figure 1.}
\label{fig:2}

\end{figure}

\begin{figure}[ht]

\includegraphics[width=1\textwidth]{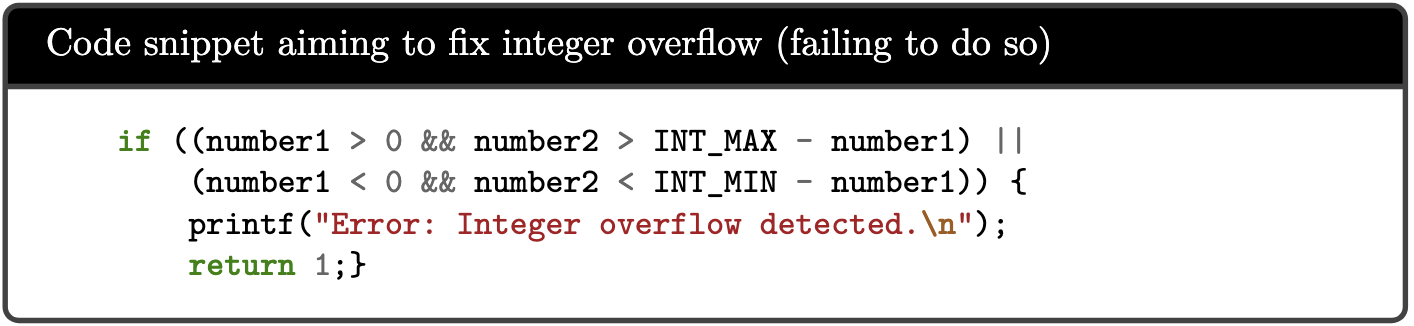}

\caption{Zero-shot prompt requesting a fix for integer overflow (failing to do so).}
\label{fig:3}
\end{figure}

\noindent Even after requesting a ``secure'' zero-shot prompt, all three original issues remain unresolved. Despite the significant advancements from GPT-3.5-turbo---which exhibited the same issue~\cite{FormAI}---to GPT-4, our motivational example indicates that GPT-4 continues to produce code with vulnerabilities. Even if specifically requested in the prompt to avoid \texttt{integer overflow} in the program, the issue persists (Figure~\ref{fig:3}).

\noindent We want to emphasize that simply requesting a secure version is not an effective approach towards achieving a secure code for the following reason: Code completion tools such as GitHub Copilot~\footnote{\url{https://github.com/features/copilot/}} or Amazon Code Whisperer~\footnote{\url{https://aws.amazon.com/codewhisperer/}} suggest code snippets based on contextual analysis and training data, which has also been shown to produce vulnerable code~\cite{nathan_nehorai_analyzing_2024}. In such scenarios, the ability to prompt is limited (it can be attempted through comments in the code). In addition, GitHub Copilot is powered by a variant of the GPT (Generative Pre-trained Transformer) model called Codex, which OpenAI developed. The underlying issue will remain if these models are not trained to produce secure code. Based on this observation, we aim to conduct a comprehensive study involving various state-of-the-art models to address our research questions.

\section{Related Work}
\label{sec:related}

This section overviews automated vulnerability detection and notable existing datasets containing vulnerable code samples for various training and benchmarking purposes.

\subsection{LLMs in Software Engineering}

In software engineering (SE), it is essential to ensure three main aspects of the code: correctness, safety, and security of the programs created. Functionally correct code should yield the expected outcomes for each input it processes. Code safety means constructing fail-safe systems, protecting against accidental or unexpected inputs that might produce logically correct but undesirable results. Software security involves fortifying the software against external threats and deliberate attacks~\cite{CordeiroFB20}. In a comprehensive study, Anwar et al.~\cite{anwar2024foundational} highlight important safety issues related to LLMs beyond SE, from the disruptive socioeconomic impacts and cybersecurity risks to ethical issues. Vassilka et al.~\cite{Vassilka} discuss the need for SE education to adapt to AI advancements and prepare future software engineers to effectively and ethically utilize these technologies in their careers.

To assess correctness, datasets such as HumanEval~\cite{chen2021codex} serve as a benchmark to measure the problem-solving abilities of AI models for problems related to language comprehension, reasoning, algorithms, simple mathematics, coding, and logical thinking.
There are several other similar datasets, such as MBPP~\cite{austin2021program} to assess code synthesis capabilities on elementary Python challenges, or CodeXGLUE~\cite{lu2021codexglue}, to test code completion, translation, and understanding of different LLMs.

Frameworks and techniques for turning prompts into executable code for SE are rapidly emerging, but the main focus mostly often functional correctness, omitting important security aspects~\cite{white_prompt_2023,yao_tree_2023,wei_chain--thought_2023}, or reliability~\cite{tihanyi2024dynamic,mirzadeh2024gsm,honarvar2023turbulence,wang2024benchmark,liang2024internal}.
There has been an arms race between researchers to excel in correctness benchmarks using zero or few-shot frameworks~\cite{guo2024deepseek,wang2023intervenor}, multi-agent frameworks~\cite{huang2023agentcoder}, fine-tuned models~\cite{muennighoff2023octopack}, and various other methods. As AI models evolve, their problem-solving capabilities improve significantly. However, whether these advancements also enhance the safety and security properties of the code they generate remains largely unclear and under-researched.

In~\cite{lin2024llm}, Lin et al. assessed different software process models to evaluate how these models affect code correctness (Pass@1\footnote{This metric highlights the model's ability to produce correct and functional code on its first try without any revisions or corrections.}). They also assessed the code quality of the AI-generated code by running static code checkers to uncover code smells\footnote{Code smells are patterns in code that hint at potential problems, making maintenance harder but not necessarily causing immediate errors. They suggest areas where the code may need to be refactored for better quality and reliability.}. This work had an interesting finding: the proposed software process models improved the quality of the generated code by significantly reducing code smells compared to what GPT-3.5-turbo outputs by itself.
Code smells or bad coding practices will not outright introduce vulnerabilities. However, several small-scale studies point to the fact that LLMs negatively impact software development from a security perspective. In~\cite{khoury_how_2023}, the authors generated $21$ small programs in five different languages: C, C++, Python, HTML, and Java. Combining manual verification with GPT-based vulnerability detection, the study found that only $5$ of the $21$ generated programs were initially secure. 

In~\cite{pearce_asleep_2021}, Pearce et al. conclude that the control group utilized GitHub's Copilot to solve arithmetic operations accurately. This work highlights an important lesson: to accurately measure the role of AI tools in code generation or completion, it is essential to choose coding scenarios mirroring a diverse set of relevant real-world settings, thereby facilitating the occurrence of various vulnerabilities. This necessitates the creation of code bases replicating a wide range of scenarios, which is one of the primary goals the FormAI dataset strives to achieve. These studies indicate that AI tools, and in particular ChatGPT, can produce code containing vulnerabilities as of today.

Ma et al.~\cite{ma_scope_2023} assessed the capabilities and limitations of ChatGPT for SE and provided initial insights into why the programs generated by language models are syntactically correct but potentially vulnerable. A study by Microsoft~\cite{imani2023mathprompter} found that GPT models encounter difficulties when accurately solving arithmetic operations. This aligns with our findings in Section~\ref{sec:motivation}.

In a comprehensive literature review, Hou et al. \cite{hou2024large} examined LLMs' application, effects, and possible limitations on SE. This study reveals that LLMs are extensively employed across software development, appearing in $229$ papers for $24$ distinct SE tasks, predominantly in code generation and program repair. It also identifies over $70$ LLMs, classifying them into three architectural types: decoder-only, encoder-decoder, and encoder-only. Each architecture serves specific functions—encoder-only for in-depth understanding, encoder-decoder for combined understanding and generation, and decoder-only primarily for generation tasks. This work highlights an interesting gap: there are dozens of research papers aiming to perform vulnerability detection in source code using machine learning (ML) and LLMs \cite{chan2023transformer,nguyen2024deep,gao2023far,gao2024learning,grishina2023earlybird,khare2023understanding,noever2023can,shestov2024finetuning,Steenhoek,sun2024llm4vuln,Tang,Thapa,zhang2023prompt},
however, assessing software safety and security properties of LLM-generated code on a large-scale has not yet been performed apart from our original work~\cite{FormAI} for C, and recently by \cite{toth2024llms} for PHP code. Both studies evaluated a single model in a zero-shot code generation scenario, while our current work also conducts a comparison of the performance of different models.

In~\cite{shumailov_curse_2023} Shumailov et al. highlighted a phenomenon known as \textit{``model collapse''}. Their research demonstrated that integrating content generated by LLMs can lead to persistent flaws in subsequent models when using the generated data for training. This hints that training ML algorithms only on purely AI-generated content is insufficient if one aims to prepare these models for detecting vulnerabilities in human-generated code. This is essentially due to using a dataset during the training phase, which is not diverse enough and misrepresents edge cases. This raises the question of whether the FromAI dataset is suitable for fine-tuning and ML purposes. It is important to note that the AI-generated code is just one part of the dataset. Most importantly, the vulnerability labeling was not done by AI but by the ESBMC formal verification tool. This way, models trained on this dataset can essentially learn the skills of a formal verification tool (or at least try to achieve the best optimal outcomes). 

The programs are generated through a dynamic zero-shot prompting method, and the generated programs are not modified by any AI system afterward. While the primary goal of our paper is to investigate and compare the secure coding abilities of different LLMs, these conditions make the \texttt{FormAI-v2} dataset suitable for ML purposes. On the other hand, AI models were trained on human-generated content; thus, the vulnerabilities produced have roots in incorrect code created by humans. Yet, as discussed in the next section, existing datasets notoriously include synthetic data (different from AI-generated), which can be useful for benchmarking vulnerability scanners but has questionable value for training purposes~\cite{chen_diversevul_2023}.

\subsection{Existing Databases for Vulnerable C Code}

We show how the \texttt{FormAI-v2} dataset compares to seven widely studied datasets containing vulnerable code and the previous version of the dataset published in~\cite{FormAI}. The examined datasets are: Big-Vul~\cite{fan_cc_2020}, Draper~\cite{russell_automated_2018,kim_draper_2018}, SARD~\cite{black_software_2018}, Juliet~\cite{jr_juliet_2012}, Devign~\cite{zhou_devign_2019}, REVEAL~\cite{chakraborty_deep_2022}, DiverseVul~\cite{chen_diversevul_2023},
and FormAI-v1~\cite{FormAI}. Table~\ref{tab:compare} presents a comprehensive comparison of the datasets across various metrics. Some of this data is derived from review papers that evaluate these datasets~\cite{jain2023code,chen_diversevul_2023}.

\begin{table*}[ht]
\centering
\caption{Comparison of Various C Code Datasets}
\label{tab:compare}
\footnotesize
\renewcommand{\arraystretch}{1.5}  
\begin{tabular}{|>{\columncolor{gray!30}\arraybackslash}m{1.4cm}|>{\centering\arraybackslash}m{0.8cm}|>{\centering\arraybackslash}m{0.8cm}|>{\centering\arraybackslash}m{0.8cm}|>{\centering\arraybackslash}m{0.8cm}|>{\centering\arraybackslash}m{0.8cm}|>{\centering\arraybackslash}m{0.8cm}|>{\centering\arraybackslash}m{0.8cm}|>{\centering\arraybackslash}m{0.7cm}|>{\centering\arraybackslash}m{1.0cm}|}
\hline

\rowcolor{gray!30}\diagbox[width=7.5em, height=3.2\line, innerleftsep=5mm, innerrightsep=1mm]{\textbf{Specs}}{\textbf{Dataset\,\,\,}}  & \texttt{Big-Vul} & \texttt{Draper} & \texttt{SARD} & \texttt{Juliet} & \texttt{Devign} & \texttt{REVEAL} & \texttt{Diverse} \texttt{Vul} & \texttt{FormAI} & \texttt{FormAI-v2} \\ 
\hline
\texttt{Language} & C/C++ & C/C++ & Multi & Multi  & C & C/C++ & C/C++ & C & C \\
\hline
\texttt{Source} & RW & Syn + RW & Syn + RW & Syn & RW & RW & RW & AI & AI \\
\hline
\texttt{Dataset size} & 189k & 1274k & 101k & 106k & 28k & 23k & 379k & 112k & 331k \\
\hline
\texttt{Vul. Snippets} & 100\% & 5.62\% & 100\% & 100\% & 46.05\% & 9.85\% & 7.02\% & 51.24\% & 62.07\% \\
\hline
\texttt{Multi. Vulns.} & x & \checkmark & x & x& x& x & x & \checkmark & \checkmark \\
\hline
\texttt{Compilable} & x & x & x & \checkmark & x& x & x & \checkmark & \checkmark \\
\hline
\texttt{Granularity} & Func & Func & Prog & Prog & Func & Func & Func & Prog & Prog \\
\hline
\texttt{Class. Type} & CVE CWE & CWE & CWE & CWE & CVE & CVE & CWE & CWE & CWE \\
\hline
\texttt{Avg. LOC.} & 30 & 29 & 114 & 125 & 112 & 32 & 44 & 79 & 86 \\
\hline
\texttt{Labelling Method} & P & S & B/S/M & B & M & P & P & FV & FV \\
\hline
\end{tabular}

\smallskip
\footnotesize
Legend:\\
\textbf{Multi}: Multi-Language Dataset, \textbf{RW}: Real World, \textbf{Syn}: Synthetic, \textbf{AI}: AI-generated, \\
\textbf{Func}: Function level granularity, \textbf{Prog}: Program level granularity, \\
\textbf{CVE}: Common Vulnerabilities and Exposures, \textbf{CWE}: Common Weakness Enumeration, \\
\textbf{P}: GitHub Commits Patching a Vulnerability, \textbf{S}: Static Analyzer, \\
\textbf{B}: By Design Vulnerable, 
\textbf{FV}: Formal Verification with ESBMC, \textbf{M}: Manual Labeling
\end{table*}

Big-Vul, Draper, Devign, REVEAL, and DiverseVul comprise vulnerable real-world functions from open-source applications. These five datasets do not include all the samples' dependencies; therefore, they are non-compilable. SARD and Juliet contain synthetic, compilable programs. In their general composition, the programs contain a vulnerable function, its equivalent patched function, and a main function calling these functions. 
All datasets indicate whether a code is vulnerable, using various vulnerability labeling methodologies such as \texttt{P}, where functions are considered vulnerable before receiving GitHub commits that address detected vulnerabilities; \texttt{M}, which involves manual labeling; \texttt{S}, which uses static analyzers; and \texttt{B}, designated as by design vulnerable without the use of a vulnerability verification tool. It's important to note that the size of these datasets can be misleading, as many include samples from languages other than the one primarily studied.
For example, SARD includes not only C and C++ but also Java, PHP, and C\#.
Moreover, newly released sets often incorporate previous datasets or scrape the same GitHub repositories, making them redundant.

For example, Draper contains C and C++ code from the SATE IV Juliet Test Suite, Debian Linux distribution, and public Git
repositories. Since the open-source functions from Debian and GitHub were not labeled, the authors used a suite of static analysis tools: CPPcheck~\cite{cppcheck} and Flawfinder~\cite{russell_automated_2018}.
However, the paper does not mention if vulnerabilities were manually verified or if any confirmation has been performed to root out false positives.
In \cite{chen_diversevul_2023}, on top of creating DiverseVul, Chen et al. merged all datasets that were based on GitHub commits and removed duplicates, thus making the most comprehensive collection of GitHub commits containing vulnerable C and C++ code.

\subsection{Vulnerability Scanning and Repair}

Software verification is crucial for ensuring software's safety and security properties. It employs a variety of techniques, each with its strengths and limitations. These techniques include manual verification, static analysis, dynamic analysis, formal verification, and increasingly, machine learning-based approaches such as those involving LLMs~\cite{cordeiro_smt-based_2012,dsilva_survey_2008,MorseCNF11,wallace_software_1989,ma_scope_2023}.
 
Manual verification involves human-driven processes such as code reviews and manual testing. While these methods effectively catch complex errors that automated tools might miss, they are labor-intensive and not scalable to large codebases or frequent updates. Static analysis evaluates source code without executing it, using static symbolic execution, data flow analysis, and control flow analysis. Style checking enforces coding standards for better readability and maintainability. 

These methods collectively enhance software integrity. The drawbacks are that this method can miss vulnerabilities that manifest only during runtime interactions and often introduce false positive results. Dynamic analysis tests the software's behavior during execution~\cite{AlshmranyABC21}. It includes fuzzing, automated testing, run-time verification, and profiling. This technique requires executable code and often significant setup to simulate different environments and may not cover all execution paths.

Formal Verification (FV) uses mathematical proofs to verify the correctness of algorithms against their specifications. It is the most rigorous form of software verification and is used in applications where reliability is critical, such as aerospace and medical devices. However, FV can be time-consuming and requires specialized knowledge, limiting its widespread adoption~\cite{CordeiroFB20}. Recent advancements include machine learning techniques, particularly LLMs, in various aspects of software verification~\cite{braberman2024tasks}. LLMs can assist in automated code review by suggesting improvements, detecting vulnerabilities, generating test cases, fixing bugs, and creating documentation. Despite their potential~\cite{hao2023v,yang2024large,quan2023xgv,10237047,10.1145/3576039,10.1145/3611643.3616256,zhang2023steam}, LLMs, on their own face limitations such as a lack of understanding of code semantics and difficulty in handling highly domain-specific knowledge~\cite{wu2023large}, and they depend heavily on the quality and variety of the training data. Using LLMs as part of a framework to complement other techniques is, however, a promising area of research~\cite{mohajer2023skipanalyzer, charalambous_new_2023, FormAI, li2023finding}. An earlier work from $2022$ examined the ability of various LLMs to fix vulnerabilities, where the models showed promising results, especially when combined. Still, the authors noted that such tools are not ready to be used in a program repair framework, where further research is necessary to incorporate bug localization. They further highlighted challenges in the tool's ability to generate functionally correct code \cite{pearce_examining_2022}. 

While LLMs struggle with detection by themselves, in~\cite{charalambous_new_2023}, the authors demonstrated that GPT-3.5-turbo could efficiently fix errors if the output of the ESBMC verifier is provided. Program repair is another emerging area where the application of LLMs is showing real promise, where in addition to fine-tuning strategies, the combination of LLMs with other tools appears to be an effective method~\cite{cao2023study,deligiannis2023fixing,fan2023automated,gao2023far,huang2023chain,islam2024code,jin2023inferfix,9809071,paul2023automated,10.1145/3597503.3608132,10.1145/3324884.3416532,Wei2023Co,widjojo2023addressing,xia2022practical,xia2023keep,10402095,10298335,zhang2023neural}.
In~\cite{wu2023effective}, the authors call for innovations to enhance automated vulnerability repair, particularly for developing more extensive training datasets to optimize LLMs.

\section{Formal Verification (FV) and Bounded Model Checking (BMC)}
\label{sec:formal}

Before presenting the methodology used to construct the dataset and examining the performance of different LLMs, this section will introduce key Formal Verification (FV) concepts to clarify the approach adopted in developing the dataset. Since manually labeling the entire dataset is not feasible for such a large volume of data, we use an FV technique known as Bounded Model Checking (BMC) to detect vulnerabilities in the generated C samples precisely. In contrast to traditional static analysis tools, which frequently produce a high number of false positives due to their reliance on pattern recognition without a solid mathematical foundation~\cite{GadelhaSC0N19}, BMC provides rigorous validation that can help minimize both false positives and false negatives in the findings.

\subsection{Preliminaries for the Data Labeling Method }
To enhance understanding and ensure the reproducibility of our methodology, we introduce some key definitions, including State Transition Systems (STS), the BMC problem, and the specific tools chosen for our labeling method, considering the many FV tools available in the market.

\subsubsection{State Transition System}

A state transition system $M = (S, T, S_0)$ represents an abstract machine consisting of a collection of states \(S\), where \(S_0 \subseteq S\) indicates the initial states, and \(T \subseteq S \times S\) specifies the transition relation, illustrating the potential state transitions within the system. Every state \(s \in S\) is characterized by the value of the program counter (\(pc\)) and the values of all program variables. The initial state \(s_0\) sets the program's starting location. Transitions between states denoted as \(T = (s_i, s_{i+1}) \in T\), between any two states \(s_i\) and \(s_{i+1}\), are associated with a logical formula \(T(s_i, s_{i+1})\) that describes the constraints on the program counter and program variables relevant to that transition.

\subsubsection{Bounded Model Checking}

BMC is employed in FV  to ascertain the correctness of a system up to a finite number of steps. This approach models the system as a finite state transition system and methodically examines its state space to a predefined depth. Recent BMC modules are capable of processing a variety of programming languages such as C, C++, JAVA, or Kotlin~\cite{sadowski2014developers, GadelhaSC0N19, white2016deep, zhao2018deepsim, CordeiroKS19, MenezesMCFC22, gadelha_esbmc_2023}. The process begins with the program code, from which a control-flow graph (CFG)~\cite{Aho:2006:CPT:1177220} is derived. In this CFG, nodes represent deterministic or nondeterministic assignments or conditional statements, while edges indicate potential changes in the program's control flow.

Essentially, each node is a block that encapsulates a set of instructions with a unique entry and exit point, and edges indicate potential transitions to other blocks. The CFG is then converted into Static Single Assignment (SSA) form and further into a State Transition System (STS), which a Satisfiability Modulo Theories (SMT) solver can interpret. The SMT solver checks if a given formula, representing the program's correctness within a bound $k$, is satisfiable, indicating the existence of a potential counterexample to the properties being verified. If no errors are found and the formula is unsatisfiable within the bound $k$, it suggests the program has no vulnerabilities within that limit. Thus, if the solver concludes satisfiability within a bound $\leq k$, it confirms the presence of a vulnerability through a counterexample.

Consider a program $\mathcal{P}$ under verification modeled as a finite STS, denoted by the triplet $\mathcal{ST} = (S, R, I)$, where $S$ represents the set of states, $R \subseteq S \times S$ represents the set of transitions, and $I \subseteq S$, including elements such as $s_n, \ldots, s_m$, marks the initial state set. In a state transition system, a state denoted as $s \in S$ consists of the program counter value, referred to as \textit{pc}, and the values of all program variables. The initial state, $s_0$, specifies the initial program location on the CFG. Every transition \(T = (s_i, s_{i+1}) \in R\), connecting two states \(s_i\) and \(s_{i+1}\), correlates with a logical expression \(T(s_i, s_{i+1})\) that constrains the program counter (\textit{pc}) and variable values pertinent to the transition. 

In the context of BMC, the properties under examination are defined as follows: $\phi(s)$ represents a logical formula reflecting states that fulfill a safety or security criterion, whereas $\psi(s)$ encodes a logical statement representing states that meet the completeness threshold, synonymous with program termination. Notably, \(\psi(s)\) incorporates loop unwinding to avoid surpassing the program's maximum loop iterations. Termination and error conditions are mutually exclusive, rendering $\phi(s) \wedge \psi(s)$ inherently unsatisfiable. If $T(s_i, s_{i+1}) \vee \phi(s)$ is unsatisfiable, state $s$ is considered a deadlock state. 

\subsubsection{The Bounded Model Checking Problem}

Based on this information, we can define the bounded model checking problem as $BMC_{\Phi}$, which involves creating a logical statement. The truth of this statement determines if the program \(\mathcal{P}\) has a counterexample with a maximum length of \(k\). The formula can only be satisfied if a counterexample fitting within the predetermined length restriction is present, i.e.:

\begin{equation}\label{eq:bmc}
BMC_{\Phi}(k) = I(s_0) \wedge \bigwedge^{k-1}_{i=1} T(s_i, s_{i+1}) \wedge \bigvee^{k}_{i=1} \neg \phi(s_i).
\end{equation}

Herein, \(I\) denotes the initial state set of \(\mathcal{ST}\), and \(T(s_i, s_{i+1})\) embodies the transition relation within \(\mathcal{ST}\) between consecutive time steps \(i\) and \(i+1\). Thus, the logical expression \(I(s_0) \wedge \bigwedge^{k-1}_{i=1} T(s_i, s_{i+1})\) depicts the execution pathways of \(\mathcal{ST}\) spanning a length \(k\), and $BMC_{\Phi}(k)$ can be satisfied if and only if for some $i \leq k$ there exists a reachable state at time step $i$ in which $\phi$ is violated. If \(BMC_{\Phi}(k)\) is satisfied, it implies a violation of \(\phi\), permitting an SMT solver to deduce a satisfying assignment from which the program variables' values can be derived to assemble a counterexample. By definition, a counterexample, or trace, for a violated property $\phi$, is defined as a finite sequence of states $s_0, \ldots, s_k$, where $s_0, \ldots, s_k \in S$ and $T(s_i, s_{i+1})$ holds for $0 \leq i < k$.
These counterexamples hold significant importance for us, as we explicitly seek out these violations to compare and determine which code generated by LLMs is ``more secure''. Fewer violated properties indicate that the LLM can produce more secure code. 

In this context, it's important to note that fewer errors in a C program generated by an LLM do not necessarily indicate superiority; the model may simply be producing simpler, shorter programs. Therefore, evaluating both property violations and code complexity metrics, such as Source Lines of Code (SLOC) or \textit{Cyclomatic Complexity} (CC) ~\cite{McCabe}, can be a good starting point to determine the complexity of the generated programs. For example, a basic ``print hello world'' program will not contain any vulnerabilities, but that doesn't mean it's a good program. This is why metrics like SLOC (Source Lines of Code) and CC (Cyclomatic Complexity) are crucial, as they help identify overly simplistic or short code that may have fewer vulnerabilities simply due to its simplicity, not because it's well-written.

If the Equation (\ref{eq:bmc}) is unsatisfiable, it implicates no error state as reachable within \(k\) steps or fewer. Hence, no software vulnerability exists within the bound \(k\). By searching for counterexamples within this bound, we can establish, based on mathematical proofs, whether a counterexample exists and whether our program $\mathcal{P}$ contains a security vulnerability. This method detects security issues such as buffer overflows, division by zero, and null dereference failures. Notably, if a program is identified as vulnerable, this determination is based on counterexamples, effectively reducing the likelihood of false positives. Conversely, in cases where no counterexample is found, we can confidently state that the program is free from vulnerabilities up to the bound $k$, thereby minimizing false negatives. By adopting this strategy, we aim to classify each program by detecting violated properties in the generated code. 

\subsection{Efficient SMT-based Context-Bounded Model Checker}
\label{esbmc}

Numerous BMC tools could meet our needs. However, we aimed to select a tool offering high performance and detection rates. Annually, the International Competition on Software Verification, known as SV-COMP, challenges various programs to detect bugs and ensure software safety. In this competition, the Efficient SMT-based Bounded Model Checker (ESBMC)~\cite{gadelha2018esbmc} stands out by solving the highest number of verification tasks within a strict 10-30 second time limit per program, as demonstrated in SV-COMP 2023~\cite{SVCOMP2023}.

Given its performance, ESBMC was selected as our primary BMC tool. As a robust, open-source model checker for C/C++, Kotlin, and Solidity programs, ESBMC addresses a wide range of safety properties and program assertions, including out-of-bounds array access, illegal pointer dereference, integer overflows, and memory leaks. It employs advanced verification techniques such as incremental BMC and \textit{k}-induction, supported by state-of-the-art SMT and Constraint Programming (CP) solvers. ESBMC's effectiveness in bug-finding is highlighted by its numerous achievements in SV-COMP, earning 6 gold, 4 silver, and 10 bronze medals.

\subsubsection{Identifiable Bugs Using ESBMC}

Although using ESBMC to identify bugs provides greater precision than traditional static analysis tools, it is also more time-consuming, requires substantial resources, and is limited to detecting a specific set of vulnerabilities. This raises a natural question: what types of vulnerabilities can BMC detect, and which are the ones it cannot?

BMCs primarily address low-level, code-centric issues such as buffer overflows, memory leaks, and assertion failures. For instance, ESBMC identifies software errors by simulating a limited portion of a program's execution with all possible inputs. However, vulnerabilities such as SQL injection, code injection, and XSS generally fall outside the scope of BMCs because creating a general mathematical model to represent how a web browser or database interprets code is highly challenging.
Additionally, SQL queries and HTML scripts can be written in various ways, making it impossible to create an exact abstract formula. This is particularly problematic because the primary goal of formal verification is to model all possible inputs to verify the system and identify property violations effectively. 

When using ESBMC, the verification result of each C program falls into one of four major categories: Verification Success ($\mathcal{VS}$), Verification Failed ($\mathcal{VF}$), Verification Unknown ($\mathcal{VU}$), and Parsing Errors ($\mathcal{ER}$), as illustrated in Table \ref{tab:four_category}. These categories are mutually exclusive, meaning a single sample cannot belong to more than one category.

\begin{table}[ht]
\centering
\begingroup
\renewcommand{\arraystretch}{1.3} 
\caption{The Four Main Categories for Vulnerability Classification With ESBMC}
\label{tab:four_category}
\begin{tabular}{@{}>{\bfseries}P{4.0cm} p{8.5cm}@{}} 
\toprule
Category & Description \\
\midrule
$\mathcal{VS}$: Verification Success & The set of samples for which the verification process was completed successfully with no vulnerabilities detected. \\
\hline
$\mathcal{VU}$: Verification Unknown \,\,\,\,\,\, (Timeout) & The set of samples for which the verification process did not complete within the allotted time frame. Although no counterexample was found within the time limit, this does not guarantee the absence of vulnerabilities in the program with a longer time frame; therefore, the verification status remains unknown.  \\
\hline
$\mathcal{VF}$: Verification Failed & The set of samples for which the verification status failed, vulnerabilities detected by ESBMC based on counterexamples. \\
\hline
$\mathcal{ER}$: Error & The set of samples for which the verification status resulted in an error. This typically occurs due to a parsing error in ESBMC, an issue in the GOTO converter, or other problems with the SMT solver. \\
\bottomrule
\end{tabular}
\endgroup
\end{table}
The most relevant category for our analysis is Verification Failed ($\mathcal{VF}$), which can be further subdivided into five main types: dereference failures ($\mathcal{DF}$), arithmetic overflows ($\mathcal{AO}$), buffer overflows ($\mathcal{BO}$), array bounds violations ($\mathcal{ABV}$), and other miscellaneous vulnerabilities ($\mathcal{MV}$). These five categories encompass 33 subcategories that ESBMC can identify, as illustrated in Table~\ref{tab:vulnerability}.

It is important to note that when we refer to ``Verification Successful'' ($\mathcal{VS}$), it indicates that the specific bugs listed in Table~\ref{tab:vulnerability} were not detected in the programs. However, this does not rule out the presence of other types of vulnerabilities in these programs, such as command injection, cryptographic weaknesses, SQL injection, and similar issues. Table~\ref{tab:vulnerability} also displays each vulnerability's corresponding Common Weakness Enumeration (CWE) number. It is important to note that ESBMC does not provide an exact mapping of vulnerabilities to CWE numbers; the mapping presented here was performed manually.

The multifaceted nature of software flaws often results in a single vulnerability associated with multiple CWE identifiers. Table \ref{tab:vulnerability} categorizes the most common vulnerabilities and the corresponding CWEs identified within these categories. In total, $42$ unique CWE were identified in the dataset.
From MITRE's Top 25 Most Dangerous Software Weaknesses for 2023 list, six is present in our list as shown in Table \ref{tab:mitre}.

\begin{table}[t!]
\centering
\caption{CWEs From 2023's MITRE Top 25.}
\begin{tabular}{cl}
\toprule
\textbf{Rank} & \textbf{CWE Description} \\
\midrule
1 & CWE-787: Out-of-bounds Write \\
4 & CWE-416: Use After Free \\
6 & CWE-20: Improper Input Validation \\
7 & CWE-125: Out-of-bounds Read \\
12 & CWE-476: NULL Pointer Dereference \\
14 & CWE-190: Integer Overflow or Wraparound \\
\bottomrule
\end{tabular}
\label{tab:mitre}
\end{table}

The remaining CWEs in the top 25 list are related to web vulnerabilities like SQL injection, XSS, and authentication, which are irrelevant to our C language samples. It is vital to emphasize that, in our situation, classifying the C programs based on CWE identifiers is not practical, contrary to what is done for other databases like Juliet. As shown in Table \ref{tab:compare}, most datasets contain only one vulnerability per sample. In the datasets ReVeal, BigVul, and Diversevul, a function is vulnerable if the vulnerability-fixing commit changes it, while in Juliet, a single vulnerability is introduced for each program.

In FormAI, a single file often contains multiple vulnerabilities. As noted, a single vulnerability can be associated with multiple CWEs. Additionally, multiple CWEs can be required for a vulnerability to be exploitable. 
As an example, \textit{``CWE-120: Buffer Copy without Checking Size of Input (Classic Buffer Overflow)''}, can happen as a result of \textit{``CWE-676: Use of Potentially Dangerous Function''}, which can be the \textit{scanf} function. If this is combined with \textit{``CWE-20: Improper Input Validation''}, it can result in \textit{``CWE-787: Out-of-bounds Write''}. Labeling the vulnerable function name, line number, and vulnerability type identified by the ESBMC module provides granular information that can benefit machine learning training. This level of detail can allow models to discern patterns and correlations with higher precision, thereby improving vulnerability prediction and detection capabilities.

\begin{table}[ht]
\centering
\caption{Detailed Categorization of Vulnerabilities Detected by ESBMC}
\begin{tabular}{>{\ttfamily}p{6.6cm} >{\ttfamily}p{1cm} >{\ttfamily}p{4cm}}
\toprule
\textbf{Description} & \textbf{CWE} & \cellcolor{gray!30}\textbf{Associated CWEs} \\
\midrule
 $\mathcal{DF}$: Dereference failures: & & \\
\quad 1. NULL pointer & CWE-476 & \cellcolor{gray!40}CWE-690, CWE-391 \\
\quad 2. Invalid pointer & CWE-822 & \cellcolor{gray!15}CWE-119, CWE-787, CWE-822 \\
\quad 3. Forgotten memory & CWE-825 & \cellcolor{gray!40}CWE-401, CWE-404, CWE-459 \\
\quad 4. Array bounds violated & CWE-125 &\cellcolor{gray!15}CWE-119, CWE-787\\
\quad 5. Invalidated dynamic object & CWE-824 & \cellcolor{gray!40}CWE-416, CWE-415 \\
\quad 6. Access to object out of bounds & CWE-125 & \cellcolor{gray!15}CWE-119, CWE-787 \\
\quad 7. Accessed expired variable pointer & CWE-416 & \cellcolor{gray!40}CWE-825 \\
\quad 8. Write access to string constant &  CWE-843 & \cellcolor{gray!15} CWE-758 \\
\quad 9. Of non-dynamic memory & CWE-590 & \cellcolor{gray!40}CWE-415, CWE-415, CWE-762 \\
\quad 10. IBTA & CWE-843 & \cellcolor{gray!15}CWE-119 \\
\quad 11. Oversized field offset & CWE-787 & \cellcolor{gray!40}CWE-119, CWE-125, CWE-823 \\
\quad 12. Data object accessed with code type & CWE-843 & \cellcolor{gray!15}CWE-686, CWE-704\\
$\mathcal{AO}$: Arithmetic overflows: & & \\
\quad 13. On sub & CWE-191 & \cellcolor{gray!40}CWE-20, CWE-190, CWE-192 \\
\quad 14. On add & CWE-190 & \cellcolor{gray!15}CWE-20, CWE-191, CWE-192 \\
\quad 15. On mul & CWE-190 & \cellcolor{gray!40}CWE-20, CWE-191, CWE-192 \\
\quad 16. Floating-point ieee\_mul & CWE-190 & \cellcolor{gray!15} CWE-681 \\ 
\quad 17. Floating-point ieee\_div & CWE-682 & \cellcolor{gray!40}CWE-369, CWE-681 \\ 
\quad 18. Floating-point ieee\_add & CWE-190 & \cellcolor{gray!15}CWE-681 \\ 
\quad 19. Floating-point ieee\_sub & CWE-190 & \cellcolor{gray!40}CWE-681 \\
\quad 20. On div & CWE-190 & \cellcolor{gray!15}CWE-20, CWE-369 \\ 
\quad 21. On shl & CWE-190 & \cellcolor{gray!40}CWE-192 \\
\quad 22. On modulus & CWE-190 & \cellcolor{gray!15}CWE-20, CWE-191 \\ 
\quad 23. On neg & CWE-191 & \cellcolor{gray!40}CWE-190, CWE-192 \\

$\mathcal{BO}$: Buffer overflow: & & \\
\quad 24. On \texttt{scanf} & CWE-120 & \cellcolor{gray!40}\multirow{1}{4cm}{\{CWE-20, CWE-121, CWE-122} \\
\quad 25. On \texttt{fscanf} & CWE-120 & \cellcolor{gray!40}CWE-129, CWE-131, CWE-628  \\
\quad 26. On \texttt{sscanf} & CWE-120 &  \cellcolor{gray!40}CWE-676, CWE-680, CWE-787\}\\
$\mathcal{ABV}$: Array bounds violations: & & \\
\quad 27. lower bound & CWE-129 &\cellcolor{gray!15}\{CWE-119, CWE-125, CWE-129\\
\quad 28. upper bound & CWE-788 & \cellcolor{gray!15} CWE-131, CWE-193, CWE-787\}\\
\quad 29. VLA array size in bytes overflows  & CWE-190 & \cellcolor{gray!40}CWE-131, CWE-680 \\

$\mathcal{MV}$: Miscellaneous Vulnerabilities: & & \\
\quad 30. Division by zero & CWE-369 & \cellcolor{gray!15}CWE-691 \\
\quad 31. The pointer to a file must be valid & CWE-476 & \cellcolor{gray!40}CWE-690, CWE-459  \\
\quad 32. Same object violation & CWE-628 & \cellcolor{gray!15}CWE-843, CWE-668  \\
\quad 33. ZOFO & CWE-761 & \cellcolor{gray!40}CWE-415, CWE-590 \\

\bottomrule
 \label{tab:vulnerability}
\end{tabular}
\footnotesize
\centering
Legend:\\
ZOFO: Operand of free must have zero pointer offset,
IBTA: Object accessed with incompatible base type
\end{table}

Since our programs exhibit numerous vulnerabilities, including multiple occurrences of the same type, categorizing each solely into one CWE group, as seen with Juliet, would be sub-optimal for training purposes. This method fails to communicate crucial details about the vulnerabilities. For instance, both ``Arithmetic overflow on add'' and ``Arithmetic overflow on div'' are assigned the same primary CWE, manifesting differently in the source code. Therefore, merely labeling them with CWEs does not offer the same level of granularity and makes the dataset less suitable for ML.

While other datasets focus more on CWEs related to vulnerabilities that could be exploited, ESBMC also detects issues related to software safety. For this reason, in the FormAI dataset, we did not assign a single CWE to each vulnerability. However, based on our mapping in Table \ref{tab:vulnerability}, one can easily associate an ESBMC vulnerability with the closest CWE number if needed.

\section{Methodology and Dataset Creation}
\label{sec:methodology}

\begin{figure}[ht!] 
\centering
\includegraphics[width=1\textwidth]{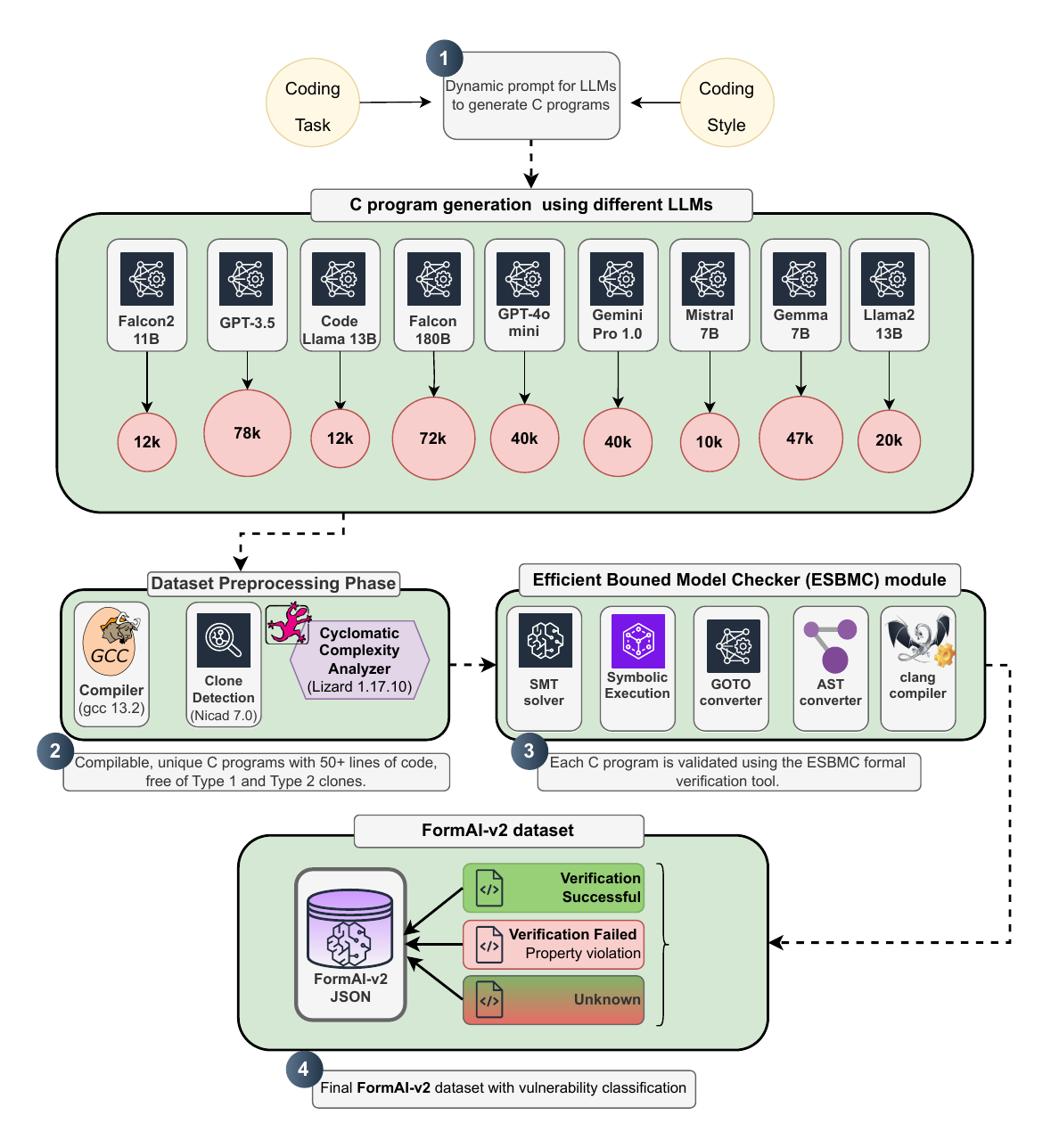}
\caption{FormAI-v2 dataset generation Framework using different LLMs.}
\label{pic:method}
\end{figure}

Figure~\ref{pic:method} provides an overview of the generation and vulnerability labeling mechanism for the FormAI-v2 dataset. This process is divided into two main components: the C program generation (consisting of 1. C program generation using different LLMs and 2. dataset preprocessing) and the classification (including 3. ESBMC classification and 4. dataset creation).

\subsection{Code Generation}

\begin{figure}[ht]
    \centering
    \includegraphics[width=1\linewidth]{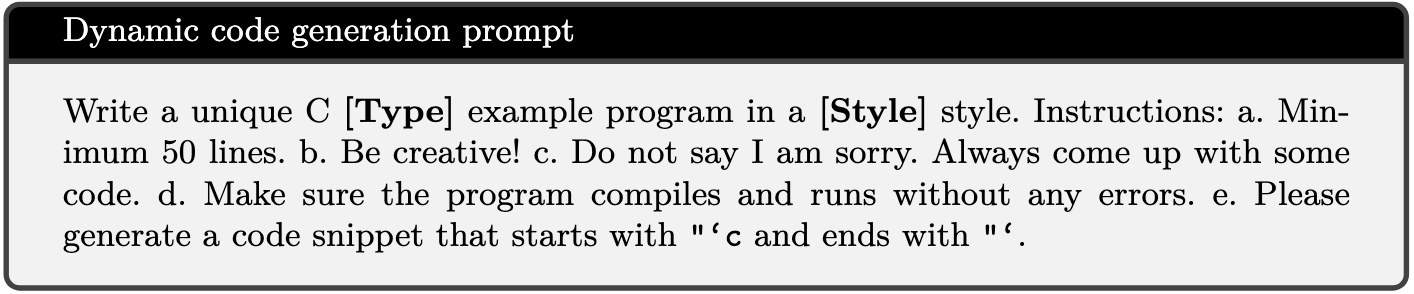}
    \caption{Dynamic Code Generation Prompt.}
    \label{fig:prompt}
\end{figure}

During the creation process, special attention was given to ensure the diversity of the \texttt{FormAI-v2} dataset, which contains $331\,000$ compilable C samples. Using a prompt like \textit{``generate a C program''} repeatedly, would yields similar outputs, such as adding two numbers or simple string manipulations, which does not satisfy our objectives. Instead, our goal is to generate a diverse and comprehensive set of small programs.
To meet this, we have developed systematic prompting method consisting a dynamic and a static part. The static component remains unchanged for all prompts, while the dynamic portion undergoes continuous variation. An example of how our prompt template looks like is shown under Figure~\ref{fig:prompt}.

The dynamic part of the prompt, highlighted as \textbf{[Type]} and \textbf{[Style]}, represent distinct categories within the prompt, each encompassing different elements. Every API call randomly selects a Type category from a set of $200$ elements. This category contains topics such as Wi-Fi Signal Strength Analyzer, QR code reader, Image Steganography, Pixel Art Generator, Scientific Calculator Implementation, etc. Similarly, a coding Style is chosen from a set of $100$ elements during each query. This helps minimize repetition, as coding styles such as ``excited'', ``relaxed'', or ``mathematical'' are randomly combined with a Type category.  Our primary objective was to identify and capture as many vulnerabilities as possible. This method can generate $200 \times 100 = 20\,000$ distinct combinations.
As demonstrated by insights from \cite{sandoval_lost_2023, pearce_examining_2022}, there's a need for a code base that supports diverse settings while ensuring tasks remain concise to fit within the token constraints of large language models (LLMs). 

This raises a key question: If we generate a dataset of over 300,000 instances but only 20,000 distinct combinations, will it lead to redundancy? Will the same or different models produce identical outputs for these repeated prompts? To address this, we will conduct clone code detection in the next section to ensure the generated code is unique. 
Selecting prompts that LLMs can efficiently process is important, therefore we designed tasks in the Type category accordingly. 
For instance, complex prompts like ``Create a CRUD application using React for the front-end, Node.js with Express for the back-end, and MongoDB for the database'' must be broken down into smaller, manageable tasks. Furthermore, tasks with different styles, such as 'File handling' with a 'romantic' versus a 'happy' style, lead to distinct outputs, which are reflected in different representations in the vector space upon tokenization. Despite potential compatibility issues between certain Type-Style combinations, encouraging LLMs to code in varied styles has generally enhanced the diversity of responses to identical Types.

Decreasing the number of unsuccessful queries by refining the prompt is important from an efficiency perspective. We have established five instructions in each prompt to minimize the error within the generated code. These instructions, along with their corresponding explanations, are the following:
\begin{enumerate}
    \item \texttt{Minimum 50 lines:} This encourages the LLM to avoid the generation of overly simplistic code with only a few lines (which occasionally still happens);
    \item \texttt{Be creative!:} The purpose of this instruction is to generate a more diverse dataset;
    \item \texttt{Do not say I am sorry:} This instruction aims to circumvent objections and responses such as ``As an AI model, I cannot generate code'', and similar statements.
    \item \texttt{Make
sure the program compiles:} This instruction encourages the model to include header files and create a complete and compilable program. 
    \item \texttt{Generate a code snippet that starts with ```c:} Enable easy extraction of the C code from the response.
\end{enumerate}
\noindent Once a C code is generated, the GNU C compiler\footnote{\url{https://gcc.gnu.org}} is employed to verify whether the corresponding code is compilable. During the code generation process, we ensure that the \texttt{FormAI-v2} dataset exclusively consists of compilable code while excluding any other code that does not meet this criterion. Different models can generate varying percentages of compilable code depending on their parameter size. Models like GPT-4o-mini, Gemini Pro, or Falcon-180B can achieve compilation rates higher than 90\%, whereas smaller models with 7B parameters typically produce C code with a compilability rate between 55-70\%. 

The primary reason for having non-compilable code was due to the absence of necessary headers, such as \texttt{math.h}, \texttt{ctype.h}, or \texttt{stdlib.h}. As the cost of generation associated with different models can significantly vary, we did not generate the same number of samples from each model. While some tested models are open source, their operational costs and GPU usage remain significant. For instance, running Falcon-180B on AWS can cost around 40 USD per hour.  Table \ref{tab:generate} presents the samples obtained from each LLM.

\begin{table*}[t]
\centering 
\scriptsize
\caption{Content of the FormAI-v2 Dataset.}
\renewcommand{\arraystretch}{1.5}
\sisetup{
    group-separator = {\,}, 
    group-minimum-digits = 4, 
    output-decimal-marker = {.}, 
    table-number-alignment = right
}
\begin{tabular}{l l l l S[table-format=5.0]} \hline
\textbf{LLM Model} & \textbf{Company} & \textbf{Size} & \textbf{License} & \textbf{Sample Size} \\\hline\hline
GPT-4o-mini & OpenAI & N/A & Proprietary & 40000 \\ \hline
Llama2 13B & Meta & 13B & Open & 20000 \\ \hline
Mistral-7B & Mistral AI & 7B & Apache 2.0 & 10000 \\ \hline
Code Llama 13B & Meta & 13B & Proprietary & 12000 \\ \hline
Gemini Pro 1.0 & Google & N/A & Proprietary & 40000 \\ \hline
Gemma-7B & Google & 7B & Gemma-TOU & 47000 \\ \hline
Falcon-180B & TII & 180B & Apache 2.0 & 72000 \\ \hline
Falcon2-11B & TII & 11B & Apache 2.0 & 12000 \\ \hline
GPT-3.5-turbo & OpenAI & 175B & Proprietary & 78000 \\ \hline
\end{tabular}
\label{tab:generate}
\end{table*}

\subsection{Clone Code Detection}
For our purposes, we do not require entirely different programs, even if they are intended to accomplish the same task. Our goal is to observe how frequently models introduce vulnerabilities into the programs. Therefore, even small changes can be interesting, or if a model repeatedly makes the same coding errors which lead to different vulnerabilities. 
Additionally, minor variations help to generalize the dataset during machine learning training, so completely removing similar code is not our objective. 

To measure how diverse the dataset is, we used a code clone detection mechanisms to flag highly similar files. Using the state-of-the-art tool NiCad 7 (Automated Detection of Near-Miss Intentional Clones)\cite{Cordy2011TheNC}, we performed clone detection within individual tasks and across datasets generated by different models. NiCad can detect four distinct types of clones, each with varying thresholds.~\texttt{Type 1} clones are exact duplicates of code fragments, with no changes except for whitespace and comments.~\texttt{Type 2} clones permit minor modifications, such as renaming variables or altering formatting, while maintaining the original logic and structure.

~\texttt{Type 3-1} clones introduce greater flexibility, allowing small additions, deletions, or modifications while preserving overall functionality.~\texttt{Type 3-2} clones allow for even more substantial changes, but the core behavior of the code remains consistent. In our clone detection process for~\texttt{Type 3-1} and~\texttt{Type 3-2} clones, we applied a 10\% threshold to eliminate similar codes. A larger threshold could eliminate valuable code fragments for machine learning, where small deviations may still result in the same vulnerability. Even slight variations can improve the training process by offering diverse representations of the same vulnerability, enabling the model to recognize it more effectively across various scenarios. Table~\ref{tab:clones} shows each dataset's number of clones identified.

\begin{table*}[ht]
\centering 
\scriptsize
\caption{Different Types of Clones Removed From the Dataset}
\renewcommand{\arraystretch}{1.5}
\sisetup{
    group-separator = {\,}, 
    group-minimum-digits = 4, 
    output-decimal-marker = {.}, 
    table-number-alignment = right
}
\begin{tabular}{l S[table-format=5.0] S[table-format=3.0] S[table-format=3.0] S[table-format=4.0] S[table-format=5.0] S[table-format=2.2]} \hline

\textbf{LLM Model} & \textbf{Sample size} & \textbf{Type1} & \textbf{Type2} & \textbf{Type 3-1} & \textbf{Type 3-2} & \textbf{$\Delta$(\%)}\\\hline\hline
Falcon2-11B & 12000 & 0 & 0 & 1 & 36 & 0.30\\ \hline
Mistral-7B & 10000 & 1 & 1 & 11 & 59 & 0.59 \\ \hline
CodeLlama-13B & 12000 & 3 & 5 & 12 & 128 & 1.10\\ \hline
GPT-3.5-turbo & 78000 & 118 & 301 & 502 & 1756 & 2.25\\ \hline
GPT-4o-mini & 40000 & 0 & 24 & 31 & 1075 & 2.69 \\ \hline
Gemini Pro 1.0 & 40000 & 12 & 150 & 187 & 1255 & 3.14\\ \hline
Falcon-180B & 72000 & 42 & 363 & 541 & 3464 & 4.81\\ \hline
Llama2-13B & 20000 & 165 & 607 & 1001 & 2214 & 11.07 \\ \hline
Gemma-7B & 47000 & 657 & 3229 & 2997 & 10199 & 21.70\\ \hline

\end{tabular}
\label{tab:clones}
\end{table*}

The last column, $\Delta(\%)$, shows the percentage of \texttt{Type 3-2} clones that were detected and removed from the dataset. A higher percentage indicates that the LLM generated more similar, redundant code samples.

In terms of clone categories, \texttt{Type 1} and \texttt{Type 2} are hierarchical: \texttt{Type 1} clones are a subset of \texttt{Type 2}, meaning that all \texttt{Type 1} clones are also considered \texttt{Type 2}. Similarly, \texttt{Type 3-1} and \texttt{Type 3-2} are inclusive, where \texttt{Type 3-1} clones fall within the broader \texttt{Type 3-2} category. In other words, $ \texttt{Type 1} \subseteq \texttt{Type 2}$ and $ \texttt{Type 3-1} \subseteq \texttt{Type 3-2}$.

However, \texttt{Type 2} is not a subset of \texttt{Type 3-1} because the threshold for \texttt{Type 2} clones is exactly zero—meaning only variable changes are allowed across the entire code with no additional modifications. In contrast, \texttt{Type 3-1} allows for up to a $10\%$ modification threshold, which can include variable changes, deletions, additions, or structural modifications, as long as they remain within the $10\%$ limit.

The most flexible clone category is \texttt{Type 3-2}, where a $10\%$ threshold applies to the entire code without restrictions. This means that any kind of modification, including variable changes throughout the entire program, is allowed. To ensure the dataset's quality, we removed all clones up to and including \texttt{Type 3-2}.

After filtering out these clones from each LLM-generated subset, we applied \texttt{Type 3-2} detection to the entire dataset to identify any similar code across different models. This process revealed an additional $283$ \texttt{Type 3-2} clones, which were subsequently removed. In total, $20\,469$ programs were excluded, resulting in a final dataset of $310\,531$ unique files. This demonstrates that the dataset is diverse, with only $6.18\%$ of the original programs being similar.

\subsection{Vulnerability Classification}

After code generation and elimination, the next step was classifying our dataset using ESBMC.  Compared to the classification in~\cite{FormAI}, a significant change has been made. 
The original work used bounded model checking (BMC)  with a bound set to \( k=1 \). For example, if a property violation occurs at level \( k=2 \), then \( BMC_{\Phi}(1) \) will not detect the vulnerability and it simply returns Verification Success, giving a false impression. As a result, in the \texttt{FormAI-v1} dataset, numerous samples were previously classified as \textit{``NON-VULNERABLE up to bound $k$''}. We have transitioned from bounded to \textit{unbounded} model checking to capture more vulnerabilities or prove their absence for each sample. This approach incrementally unwinds the program until a bug is found or the completeness threshold is reached, meaning all possible terminating states have been explored. Incremental BMC ensures that smaller problems are solved sequentially, avoiding the need to guess an upper bound for verification. 

Applying these settings, we have successfully identified more vulnerabilities in the programs. Consequently, if the verification process is completed successfully, we can conclude that the program has no violated properties (that can be detected by the currently used ESBMC version). While this approach requires significantly more computational power, it has proven effective in revealing more vulnerabilities or proving their absence, as we will demonstrate in Section~\ref{sec:Discussion}.

\subsubsection{ESBMC Parameter Selection}

Model-checking tools like ESBMC provide various parameters, and the identified vulnerabilities may differ depending on the parameters chosen.  Default parameters, such as those used in competitions like SV-COMP, may not be suitable for all software types, potentially leading to fewer detected vulnerabilities. This naturally leads to the question: which options should we use? Should we use the \texttt{k-induction} switch with a large time limit, such as 100 seconds, or should we opt for the \texttt{bmc} switch? These questions are not straightforward. To address them, we conducted a detailed analysis to understand how different settings impact verification outcomes. We have randomly selected $1\,000$ samples from the dataset, serving as the basis for selecting the ESBMC parameters for the entire dataset.  By experimenting with different switches and time frames, we were able to select the options that best met our needs.

For these samples, we tested multiple parameter configurations of ESBMC to determine which settings yielded the best results regarding runtime efficiency and vulnerability detection.
We focus on two objectives. Firstly, to minimize verification unknown outcomes ($\mathcal{VU}$) through the $t$ (time) parameter and preferably completing the verification process; and secondly, to identify as many vulnerabilities as possible. 
Table~\ref{tab:ESBMC_parameters} illustrates the verification outcomes of the 1,000 samples, demonstrating how various combinations of unwind (u) and time (t), alongside the utilization of \textit{k}-induction, incremental BMC, or falsification techniques, impact the results.

\begin{table}[hbpt]
\centering
\caption{Classification Results for the 1000-Sample Dataset With Varying Parameters.}
\renewcommand{\arraystretch}{1.5}
\setlength{\tabcolsep}{6pt}
\sisetup{
    group-separator = {\,}, 
    group-minimum-digits = 4, 
    output-decimal-marker = {.}, 
    table-number-alignment = right
}
\begin{tabular}{l l l l l | l S[table-format=5.0] S[table-format=5.0] S[table-format=5.0] S[table-format=5.0] S[table-format=5.0]} \hline
\multicolumn{5}{c|}{\cellcolor{gray!25}\textbf{ESBMC Parameters}} & \multicolumn{6}{c}{\cellcolor{gray!25}\textbf{RESULTS}} \\ \hline

\textbf{u} & \textbf{time} & \textbf{k-ind} & \textbf{bmc} & \textbf{fls} & \textbf{Runtime} & \textbf{$|\phi|$} & \textbf{$\mathcal{VS}$} & \textbf{$\mathcal{VF}$} & \textbf{$\mathcal{VU}$} & \textbf{$\mathcal{ER}$} \\ \hline\hline
x & 300 & \cellcolor{green!25}\checkmark & \cellcolor{red!25}x & \cellcolor{red!25}x & 1698:53 & 1678 & 471 & 491 & 25 & 13 \\ \hline
2 & 1000 & \cellcolor{red!25}x & \cellcolor{red!25}x & \cellcolor{red!25}x & 1418:03 & 1638 & 505 & 407 & 70 & 18 \\ \hline
3 & 1000 & \cellcolor{red!25}x & \cellcolor{red!25}x & \cellcolor{red!25}x & 2100:36 & 1620 & 495 & 390 & 94 & 21 \\ \hline
x & 100 & \cellcolor{green!25}\checkmark & \cellcolor{red!25}x & \cellcolor{red!25}x & 653:05 & 1583 & 486 & 468 & 33 & 13 \\ \hline
2 & 100 & \cellcolor{red!25}x & \cellcolor{red!25}x & \cellcolor{red!25}x & 224:25 & 1580 & 496 & 393 & 96 & 15 \\ \hline
1 & 1000 & \cellcolor{red!25}x & \cellcolor{red!25}x & \cellcolor{red!25}x & 419:45 & 1529 & 538 & 428 & 21 & 13 \\ \hline
x & 30 & \cellcolor{green!25}\checkmark & \cellcolor{red!25}x & \cellcolor{red!25}x & 216:28 & 1513 & 494 & 448 & 45 & 13 \\ \hline
x & 30 & \cellcolor{red!25}x & \cellcolor{green!25}\checkmark & \cellcolor{red!25}x & 216:20 & 1511 & 494 & 448 & 45 & 13 \\ \hline
x & 30 & \cellcolor{red!25}x & \cellcolor{red!25}x & \cellcolor{green!25}\checkmark & 232:36 & 1511 & 494 & 448 & 45 & 13 \\ \hline
2 & 30 & \cellcolor{red!25}x & \cellcolor{red!25}x & \cellcolor{red!25}x & 99:05 & 1500 & 486 & 371 & 129 & 14 \\ \hline
1 & 100 & \cellcolor{red!25}x & \cellcolor{red!25}x & \cellcolor{red!25}x & 79:09 & 1465 & 536 & 421 & 30 & 13 \\ \hline
x & 10 & \cellcolor{green!25}\checkmark & \cellcolor{red!25}x & \cellcolor{red!25}x & 84:11 & 1432 & 500 & 430 & 57 & 13 \\ \hline
3 & 100 & \cellcolor{red!25}x & \cellcolor{red!25}x & \cellcolor{red!25}x & 344:01 & 1408 & 478 & 350 & 158 & 14 \\ \hline
1 & 10 & \cellcolor{red!25}x & \cellcolor{red!25}x & \cellcolor{red!25}x & 21:47 & 1351 & 527 & 399 & 61 & 13 \\ \hline
2 & 10 & \cellcolor{red!25}x & \cellcolor{red!25}x & \cellcolor{red!25}x & 47:48 & 1272 & 469 & 330 & 187 & 14 \\ \hline
3 & 10 & \cellcolor{red!25}x & \cellcolor{red!25}x & \cellcolor{red!25}x & 62:14 & 951 & 433 & 272 & 281 & 14 \\ \hline
x & 1 & \cellcolor{green!25}\checkmark & \cellcolor{red!25}x & \cellcolor{red!25}x & 13:23 & 941 & 474 & 336 & 177 & 13 \\ \hline
x & 1 & \cellcolor{red!25}x & \cellcolor{red!25}x & \cellcolor{green!25}\checkmark & 13:39 & 938 & 475 & 335 & 177 & 13 \\ \hline
x & 1 & \cellcolor{red!25}x & \cellcolor{green!25}\checkmark & \cellcolor{red!25}x & 13:34 & 936 & 476 & 334 & 177 & 13 \\ \hline
1 & 1 & \cellcolor{red!25}x & \cellcolor{red!25}x & \cellcolor{red!25}x & 7:41 & 911 & 487 & 323 & 177 & 13 \\ \hline
2 & 1 & \cellcolor{red!25}x & \cellcolor{red!25}x & \cellcolor{red!25}x & 10:30 & 559 & 404 & 205 & 377 & 14 \\ \hline
10 & 1 & \cellcolor{red!25}x & \cellcolor{red!25}x & \cellcolor{red!25}x & 14:14 & 158 & 224 & 79 & 684 & 13 \\ \hline
x & 10 & \cellcolor{red!25}x & \cellcolor{red!25}x & \cellcolor{red!25}x & 152:41 & 69 & 75 & 25 & 887 & 13 \\ \hline

\end{tabular}
\vspace{5pt}
\center
\textbf{Legend:} \\
\checkmark: Enabled; 
x: Not set; 
$|\phi|$:  Number of Vulnerabilities detected; \textbf{k-ind:} k-induction; \\\textbf{bmc:} incremental-bmc;  \textbf{fls:} falsification technique; \textbf{u}: unwind; (Runtime in (m:s))     
\label{tab:ESBMC_parameters}
\end{table}

\begin{figure}[hbpt]
    \centering
    \includegraphics[width=1\linewidth]{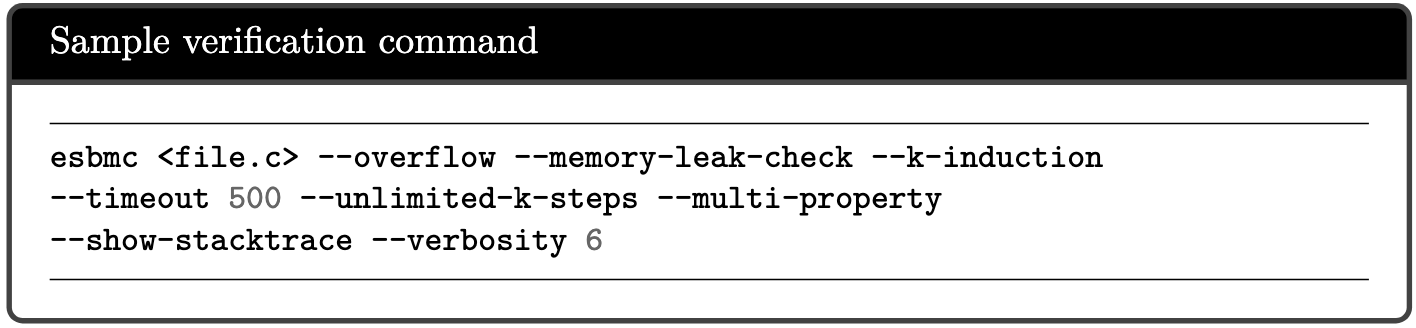}
    \caption{ESBMC Command Employed to Verify Each Sample in the Dataset.}
    \label{list:command}
\end{figure}

In this context, ``unwind'' refers to the number of iterations for which we should unroll the loops. For example, u = 1 means that we unroll a loop for only one iteration, while u = $\infty$ indicates no limit on the number of iterations for loop unrolling. Our analysis revealed that merely increasing the unwind parameter $u$ while keeping a short timeout (e.g., 1 second) often leads to timeouts. For example, setting the unwind to $10$ with a 1-second timeout resulted in most samples (684) falling into $\mathcal{VU}$. A larger unwind parameter enhances the detection of vulnerabilities in loops, provided there is sufficient processing time. We can also observe that the \texttt{k-induction} switch increases the number of detected vulnerabilities, $|\phi|$, as the allotted time increases. Therefore, the best approach for us is to set the timeout as high as possible with \texttt{k-induction} enabled. 
In our architecture, we set the timeout to 500 seconds and allowed unlimited k-steps, transitioning from bounded to \textit{unbounded model} checking. This adjustment ensures that if the verification is completed within this time frame, we either identify a counterexample or confirm the absence of the examined vulnerabilities.
For our dataset classification, we used a machine with 192 CPUs and 1.5 TB of memory, which allowed us to set a time frame of 500. A larger time frame is not feasible in our test environment, as concurrently verifying 192 C programs would exceed the 1.5 TB memory capacity. Based on our experiments, we used the following ESBMC switches during our experiments, as depicted in Figure~\ref{list:command}.

Note, that the \texttt{--overflow}, \texttt{--memory-leak-check}, and \texttt{--multi-property} switches are used to identify the maximum number of potential vulnerabilities. These switches do not affect the running time. Using these parameters on our $1000$ sample set, 416 files were deemed non-vulnerable, while 519 files were vulnerable. Among these 519 files, a total of 2116 unique vulnerabilities were detected. Considering the classification of $331\,000$ programs, the worst-case scenario is that every program from FormAI-v2 would utilize its allocated time, resulting in 500 seconds dedicated to verifying each sample. Using 192 CPU threads, the entire verification process on our experimental setup would take approximately $9,97$ days in this worst-case scenario, calculated as $331\,000 \times 500 / 60 / 60/ 24 / 192$.

\section{Verification Results}\label{sec:Discussion}

In this section, we summarize our key results, beginning with an analysis of statistics for the entire dataset, focusing on overall verification outcomes and vulnerability types. It is important to note that the analysis is based on $310,531$ programs, as all clones up to \texttt{Type 3-2} have been excluded from the initial $331,000$. We then evaluate each LLM, comparing the complexity of the code they generate and their secure coding capabilities. This is followed by evaluating each LLM and comparing the complexity of the code they generate and their secure coding capabilities.

In the original FormAI dataset, only $112,000$ compilable C samples were created using GPT-3.5-turbo. Furthermore, the complexity of each program was not measured. This research closes this gap by comparing nine state-of-the-art LLMs and providing a vulnerability-labelled dataset to the research community. We have examined  $26\,633\,156$ lines of C code, with an average of $85.77$ lines per sample. 
In total, we performed the verification process on $310\,531$ C program files, and our results for the entire dataset are shown in Table \ref{tab:summary}. The TOP 10 violations throughout the entire dataset are presented in Table~\ref{tab:totalviolation}. Table \ref{tab:bigvul} provides a breakdown of the distribution for each of the top five main categories of vulnerabilities.

\begin{table}[htbp]
\centering
\scriptsize
\renewcommand{\arraystretch}{1.5}
\caption{Overview of Statistics and Verification Results for Each LLM.}
\begin{tabular}{@{}lrrrrrrrrr@{}} 
\toprule
\begin{tabular}[c]{@{}c@{}}\textbf{Model}\\ \textbf{Name}\end{tabular} & \begin{tabular}[c]{@{}c@{}}\textbf{Samples}\\ \textbf{w.o clones}\end{tabular} & \begin{tabular}[c]{@{}c@{}}\textbf{Max}\\ \textbf{SLOC}\end{tabular} & \begin{tabular}[c]{@{}c@{}}\textbf{Avg}\\ \textbf{SLOC}\end{tabular} & \begin{tabular}[c]{@{}c@{}}\textbf{Avg}\\ \textbf{CC}\end{tabular} & \begin{tabular}[c]{@{}c@{}}$\bm{\mathcal{VS}}$\\  \textbf{(\%)} \end{tabular} & \begin{tabular}[c]{@{}c@{}}$\bm{\mathcal{VU}}$\\  \textbf{(\%)}\end{tabular} & \begin{tabular}[c]{@{}c@{}}$\bm{\mathcal{VF}}$\\ \textbf{(\%)}\end{tabular} & \begin{tabular}[c]{@{}c@{}}$\bm{\mathcal{ER}}$\\  \textbf{(\%)}\end{tabular} \\
\midrule
Gemma-7B & $36\,787$ & $351$ & $67.28$ & $5.25$ & $11.62$ & $16.30$ & $67.01$ & $5.07$ \\
GPT-3.5-turbo & $76\,168$ & $616$ & $96.79$ & $6.07$ & $7.29$ & $26.09$ & $65.07$ & $1.55$ \\
Gemini Pro 1.0 & $38\,695$ & $332$ & $98.87$ & $4.56$ & $9.49$ & $24.13$ & $63.91$ & $2.47$ \\
Falcon2-11B & $11\,946$ & $338$ & $77.75$ & $6.34$ & $10.28$ & $24.56$ & $63.16$ & $2.00$ \\
Mistral-7B & $9\,934$ & $161$ & $75.06$ & $3.84$ & $8.36$ & $25.88$ & $62.08$ & $3.68$ \\
Falcon-180B & $68\,463$ & $181$ & $71.93$ & $4.38$ & $6.48$ & $28.67$ & $62.07$ & $2.78$ \\
GPT-4o-mini & $38\,921$ & $347$ & $103.58$ & $3.40$ & $4.23$ & $36.77$ & $57.14$ & $1.86$ \\

CodeLlama-13B & $11\,838$ & $258$ & $83.36$ & $4.54$ & $15.48$ & $29.52$ & $52.71$ & $2.39$ \\

Llama2-13B & $17\,779$ & $207$ & $75.51$ & $4.12$ & $12.36$ & $31.78$ & $51.30$ & $4.56$ \\

\bottomrule
\rowcolor{gray!30} \textbf{FormAI-v2}  & \bm{$310\,531$} & \bm{$616$} & \bm{$85.77$} & \bm{$4.85$} & \bm{$8.27$} & \bm{$26.99$} & \bm{$62.07$} & \bm{$2.67$} \\
\end{tabular}
\label{tab:summary}
\centering
Legend:\\

\textbf{Max SLOC}: Maximum Source Lines of Code in a sample.
\textbf{Avg SLOC}: Average Source Lines of Code per sample.
\textbf{Avg CC}: Average Cyclomatic Complexity, which measures the complexity of the code based on the number of linearly independent paths. 
\textbf{$\mathcal{VS}$}: Verifications Success.
\textbf{$\mathcal{VU}$}: Verification Unkown.
\textbf{$\mathcal{VF}$}: Verification Failed (vulnerable).
\textbf{$\mathcal{ER}$}: Error.
\end{table}

\begin{table}[htbp]
\centering
\scriptsize
\caption{Top 10 Violations Across All Categories in FormAI-v2 dataset.}
\begin{tabular}{cllrr}
\toprule
\textbf{Rank} & \textbf{Category} & \textbf{Violation Type} & \textbf{Count} & \cellcolor{gray!30}\textbf{Percentage} \\
\midrule
1 & $\mathcal{DF}$ & Dereference failure: NULL pointer & $289\,548$ & \cellcolor{gray!30}37.83\% \\
2 & $\mathcal{BO}$ & Buffer overflow on \texttt{scanf} & $214\,255$ & \cellcolor{gray!30}27.99\% \\
3 & $\mathcal{DF}$ & Dereference failure: invalid pointer &  $73\,838$ & \cellcolor{gray!30}9.65\% \\

4 & $\mathcal{DF}$ & Dereference failure: array bounds violated & $23\,586$ & \cellcolor{gray!30}3.08\% \\
5 & $\mathcal{ABV}$ & Array bounds violated: upper bound & $23\,380$ & \cellcolor{gray!30}3.05\% \\
6 & $\mathcal{DF}$ & Dereference failure: forgotten memory &  $21\,108$& \cellcolor{gray!30}2.76\% \\
7 & $\mathcal{ABV}$ & Array bounds violated: lower bound & $19\,918$ & \cellcolor{gray!30}2.60\% \\
8 & $\mathcal{AO}$ & Arithmetic overflow on sub & $18\,345$ & \cellcolor{gray!30}2.40\% \\
9 & $\mathcal{AO}$ & Arithmetic overflow on add & $15\,966$ & \cellcolor{gray!30}2.09\% \\
10 & $\mathcal{AO}$ & Arithmetic overflow on mul & $12\,462$ & \cellcolor{gray!30}1.63\% \\
\bottomrule
\end{tabular}
\label{tab:totalviolation}
\end{table}

\begin{table}[t]
\centering
\scriptsize
\caption{Detailed Categorisation of Vulnerabilities in the Entire Dataset}
\sisetup{
    group-separator = {\,},
    group-minimum-digits = 4,
    output-decimal-marker = {.},
    table-number-alignment = right
}
\begin{tabular}{p{8cm} S[table-format=6.0] S[table-format=2.2, table-space-text-post=\%]}
\toprule
\textbf{Description} & \textbf{Count} & \cellcolor{gray!30}\textbf{Percentage} \\
\midrule
\textbf{Dereference failures}: & & \\
\quad - NULL pointer & 289548 & \cellcolor{gray!30}37.83\% \\
\quad - Invalid pointer & 73838 & \cellcolor{gray!30}9.65\% \\
\quad - Forgotten memory & 21108 & \cellcolor{gray!30}2.76\% \\
\quad - Array bounds violated & 23586 & \cellcolor{gray!30}3.08\% \\
\quad - Invalidated dynamic object & 3145 & \cellcolor{gray!30}0.41\% \\
\quad - Access to object out of bounds & 3221 & \cellcolor{gray!30}0.42\% \\
\quad - Accessed expired variable pointer & 1227 & \cellcolor{gray!30}0.16\% \\
\quad - Write access to string constant & 913 & \cellcolor{gray!30}0.12\% \\
\quad - Non-dynamic memory & 342 & \cellcolor{gray!30}0.04\% \\
\quad - Object accessed with incompatible base type & 379 & \cellcolor{gray!30}0.05\% \\
\quad - Oversized field offset & 170 & \cellcolor{gray!30}0.02\% \\
\quad - Data object accessed with code type & 14 & \cellcolor{gray!30}0.00\% \\
\textbf{Arithmetic overflows}: & & \\
\quad - On sub & 18345 & \cellcolor{gray!30}2.40\% \\
\quad - On add & 15966 & \cellcolor{gray!30}2.09\% \\
\quad - On mul & 12462 & \cellcolor{gray!30}1.63\% \\
\quad - IEEE mul & 9673 & \cellcolor{gray!30}1.26\% \\
\quad - IEEE div & 3522 & \cellcolor{gray!30}0.46\% \\
\quad - IEEE add & 2375 & \cellcolor{gray!30}0.31\% \\
\quad - IEEE sub & 1632 & \cellcolor{gray!30}0.21\% \\
\quad - On div & 813 & \cellcolor{gray!30}0.11\% \\
\quad - On shl & 972 & \cellcolor{gray!30}0.13\% \\
\quad - On modulus & 348 & \cellcolor{gray!30}0.05\% \\
\quad - On neg & 155 & \cellcolor{gray!30}0.02\% \\
\textbf{Buffer overflows}: & & \\
\quad - On \texttt{scanf} & 214255 & \cellcolor{gray!30}27.99\% \\
\quad - On \texttt{fscanf} & 8252 & \cellcolor{gray!30}1.08\% \\
\quad - On \texttt{sscanf} & 4184 & \cellcolor{gray!30}0.55\% \\
\textbf{Array bounds violations}: & & \\
\quad - Upper bound & 23380 & \cellcolor{gray!30}3.05\% \\
\quad - Lower bound & 19918 & \cellcolor{gray!30}2.60\% \\
\quad - VLA array size in bytes overflows address space size & 4222 & \cellcolor{gray!30}0.55\% \\
\textbf{Miscellaneous Vulnerabilities}: & & \\
\quad - Division by zero & 4311 & \cellcolor{gray!30}0.56\% \\
\quad - The pointer to a file object must be a valid argument & 1225 & \cellcolor{gray!30}0.16\% \\
\quad - Invalid Function argument issues & 443 & \cellcolor{gray!30}0.06\% \\
\quad - Same object violation & 123 & \cellcolor{gray!30}0.02\% \\
\quad - Operand of free must have zero pointer offset & 134 & \cellcolor{gray!30}0.02\% \\
\bottomrule
\end{tabular}
\label{tab:bigvul}
\end{table}

During the $500$-second verification time-frame, ESBMC identified $192\,757$ unique programs with vulnerabilities. In contrast, only $25\,674$ programs, representing $8.27\%$, were verified as secure.
Expanding computational resources may increase the number of programs uncovered from $\mathcal{VU}$, thereby potentially extending the $\mathcal{VF}$ category.
These results provide an even better lower bound compared to~\cite{FormAI}, on what percentage of LLM-generated code is vulnerable. The situation is more concerning than merely stating that 62.07\% of the generated files are vulnerable, as a single file can contain multiple vulnerabilities. On average, each file contains $3.97$ vulnerabilities. The total number of property violations detected by ESBMC for the overall dataset is $765\,366$.

The most common type of vulnerability is related to ``Dereference failures'' accounting for $54.54\%$ of the cases, predominantly due to NULL pointer issues. This category includes a variety of pointer-related issues, such as invalid pointers, a forgotten memory, and array-bounds violations, among others.
``Buffer overflows'', mainly triggered by the scanf function, comprise a significant  $27.99$\% of the vulnerabilities. This highlights common issues in handling buffer sizes and input functions.
``Arithmetic overflows'' are also notable, covering various operations like subtraction, addition, multiplication, and division, indicating frequent issues in handling numeric calculations without adequate checks.
The table further lists ``Array bounds violations'' and ``Division by zero'' as common issues, illustrating challenges in correctly managing arrays and arithmetic operations.
A smaller portion of the table covers ``Miscellaneous Vulnerabilities'' which includes a variety of less frequent but notable issues such as invalid file object pointers and operand violations in memory deallocation.
Overall, the data emphasizes the need for robust handling of pointers, buffers, and numeric operations within the source code to mitigate the risk of vulnerabilities.

\subsection{General observation about code complexity}
\begin{figure}[b]
    \centering
    \begin{subfigure}[b]{0.46\textwidth}
        \includegraphics[width=\textwidth, trim=0.8cm 0.8cm 0.8cm 0cm, clip]{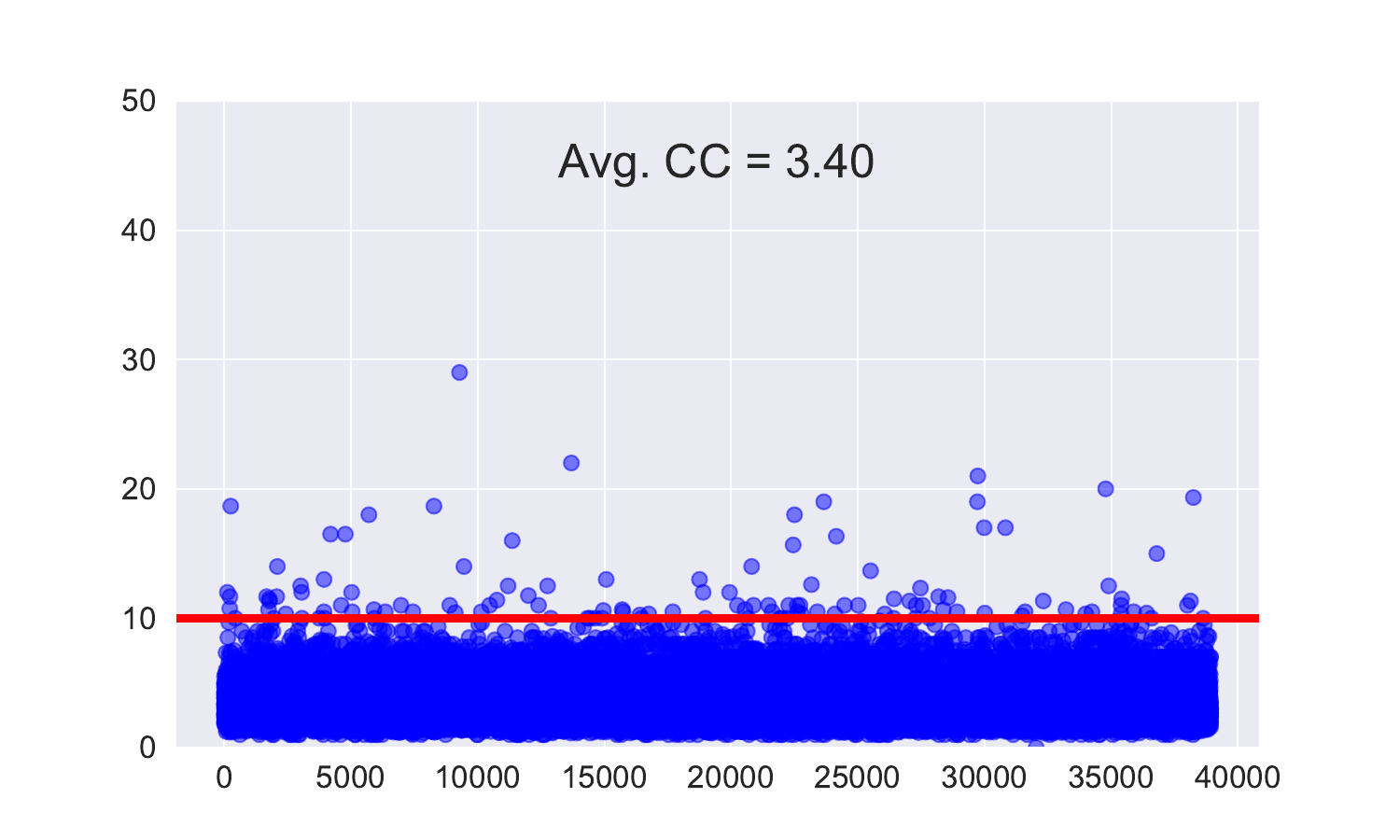}
        \caption{GPT4o-mini.}
    \end{subfigure}
    \hfill 
    \begin{subfigure}[b]{0.46\textwidth}
        \includegraphics[width=\textwidth, trim=0.8cm 0.8cm 0.8cm 0cm, clip]{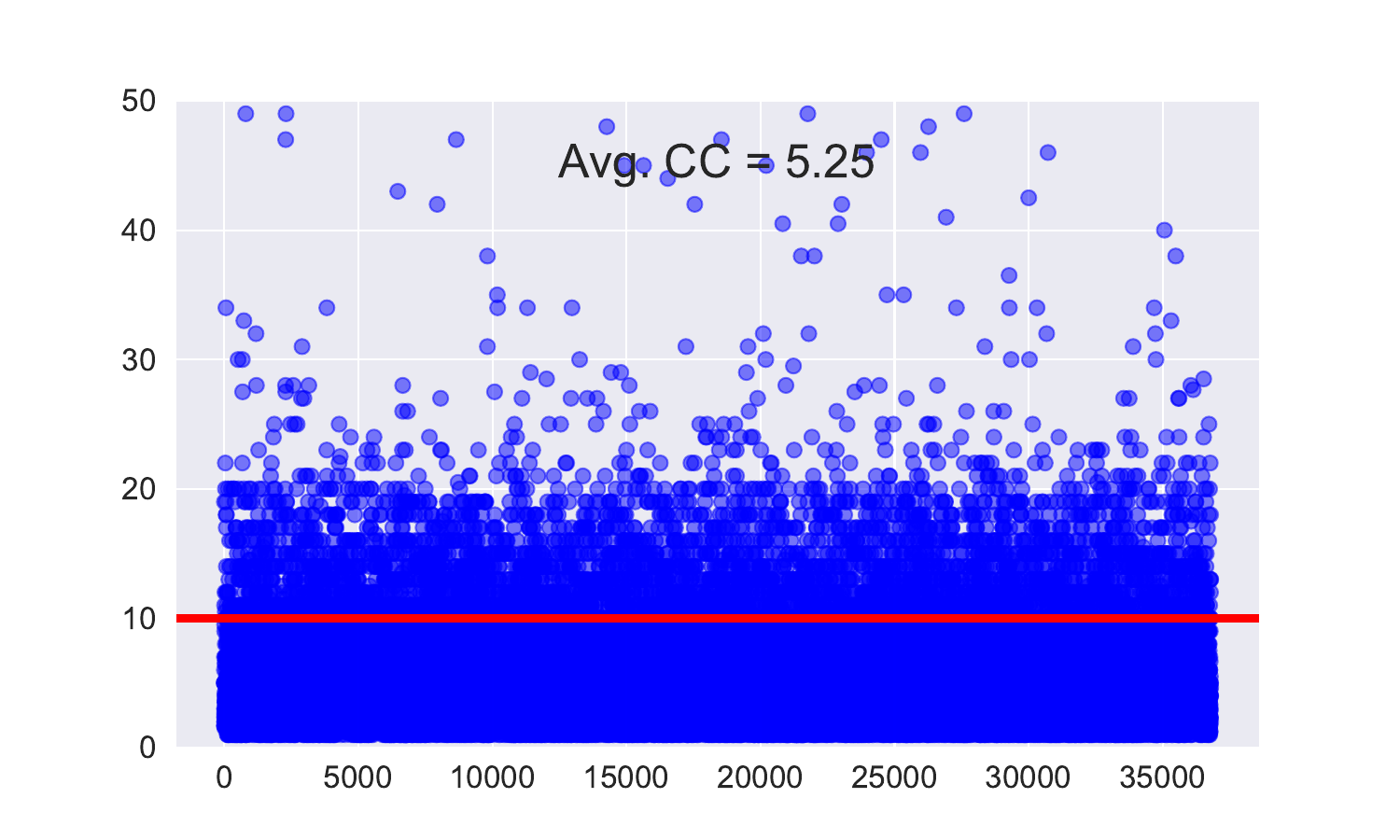}
        \caption{Gemma-7B.}
    \end{subfigure}
    \caption{Comparison of Cyclomatic Complexity between GPT4o-mini and Gemma-7B.}
    \label{fig:cyclomatic_complexity}
\end{figure}

NIST defines Cyclomatic Complexity (CC) as ``the amount of decision logic in a source code function'' and recommends a maximum value of 10~\cite{mikejo5000_code_2024}. According to NIST, ``higher numbers are bad and lower numbers are good.'' As Figure~\ref{fig:cyclomatic_complexity} shows, many individual programs generated by Gemma-7B exceed the threshold of 10. While SLOC and CC cannot be used to determine whether code is vulnerable directly, we observed that higher cyclomatic complexity can lead to an increased likelihood of vulnerabilities. Models such as GPT-3.5-turbo, Gemma-7B, and Falcon2-11B, which have high CC, also display the highest rates of verification failures.

As earlier shown in Table \ref{tab:summary}, the Avg. CC (Average cyclomatic complexity per sample)  and Avg. SLOC (Average Source
Lines of Code per sample) provide insight into the complexity of the code generated by a certain model. As previously mentioned, if a model produces only non-vulnerable code, it doesn't necessarily indicate high quality; it could suggest that the generated code is very simple (e.g., generating only ``print 'hello world''' examples). While observing SLOC and CC cannot precisely determine a model's code quality, it is interesting to observe that GPT-4o-mini, CodeLlama-13B, and Llama2-13B had the least lowest verification failed results and the lowest CC scores.

The analysis of Table \ref{tab:summary} shows that GPT-4o-mini does not necessarily generate shorter or simpler code. It produces the longest C programs, with an average SLOC of $103.48$, and has the highest verification unknown score ($36.77$\%), indicating that the ESBMC verification process takes longer for GPT-4o-mini samples. In contrast, Gemma-7B generates the shortest average SLOC and also has the lowest verification unknown result ($16.30$\%). Additionally, GPT-4o-mini produces code with a lower CC, which implies better maintainability and quality, while Gemma-7B has a much higher average CC.

\subsection{Keyword Frequency}
\begin{figure}[ht]
    \centering
    
        \includegraphics[width=1\linewidth]{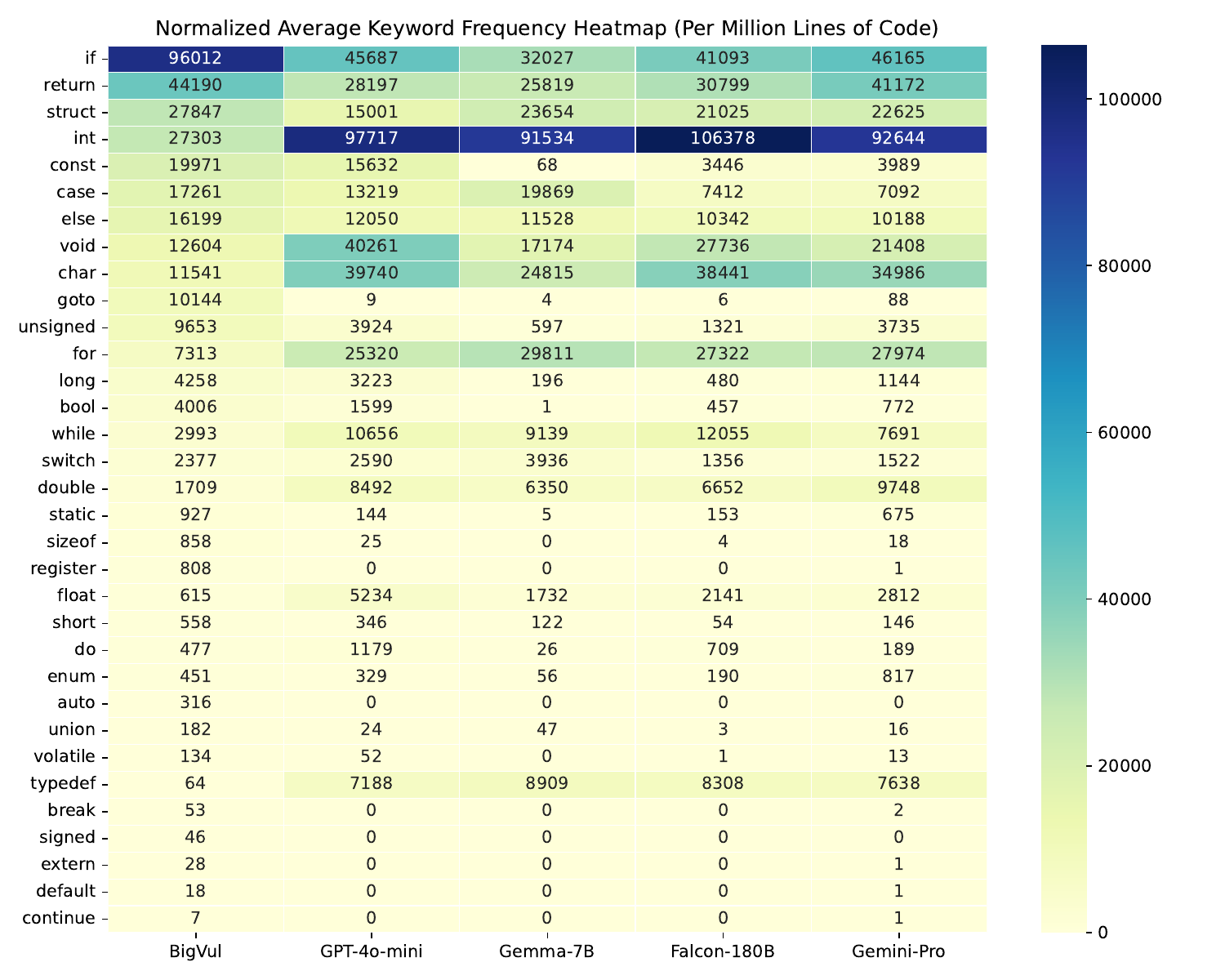}

    \caption{32 C keyword distribution}
    \label{fig:statistics}
\end{figure}
When assessing vulnerabilities in LLM-generated code, a key question arises: How does LLM-generated code compare to human-written code? If the generated code differs significantly, the dataset may not support meaningful comparisons with real-world code. A practical starting point for this analysis is comparing keyword frequencies. In real-world C/C++ projects, such as those from GitHub and datasets like BigVul, common keywords include ‘if’, ‘return’, ‘struct’, ‘int’, and ‘const’, while less frequent keywords include ‘continue’, ‘default’, and ‘extern’. Significant differences in keyword frequencies between LLM-generated and real-world code would question the dataset’s validity.

To investigate, we used a token-based keyword-counting method to analyze the frequency of 32 C keywords in each LLM-generated subset. Ideally, LLM-generated code should exhibit a similar keyword distribution to real-world code. Figure \ref{fig:statistics} shows the normalized keyword frequency (occurrences per million lines of code) for various LLM-generated codes, with BigVul as a real-world benchmark. The heatmap reveals that LLM-generated and real-world code have closely matching keyword distributions, likely due to the LLMs being trained on human-written GitHub projects.

While there are slight variations in the distribution between LLMs and BigVul mainly for the less frequent words, LLMs show great similarity on how they handle statements, expressions, and variables in distinct ways. Note, that while all LLM generated codes are fully compilable on our dataset, this is not the case with BigVul samples and other human written code datasets.

\subsection{Vulnerability Ranking}

Table~\ref{table:11a} (Parts I, II, and III) provides an overview of the top 10 vulnerabilities generated by each model. Note that raw vulnerability counts are not directly comparable due to the differing number of samples produced by each model. To enable a fair comparison across LLMs, the table also includes the percentage representation of each vulnerability.
This analysis does not offer a comprehensive review of all identified CWEs but focuses on vulnerabilities explicitly verified by ESBMC.

Buffer overflow vulnerabilities related to \texttt{scanf} are consistently ranked among the top three across all LLM models. The functions \texttt{scanf}, \texttt{fscanf}, and \texttt{sscanf} do not restrict the input size of their respective buffers, creating a risk of buffer overflow. This vulnerability can allow attackers to execute arbitrary code or trigger crashes. As previously mentioned, these issues relate to several CWEs, including CWE-676, CWE-20, and CWE-787. Although buffer overflow is a type of out-of-bounds write, CWE-787 covers a broader range of vulnerabilities. CWE-120 specifically addresses classic buffer overflow scenarios caused by unchecked input sizes during buffer copy operations.
While more complex issues like arithmetic overflows and array bounds violations require deeper programming context, simpler issues such as \texttt{scanf} errors should be easier to avoid. However, all tested models consistently exhibit buffer overflow errors with \texttt{scanf}.

\begin{table}[htbp]
\centering
\caption{Top 10 Vulnerabilities in LLM Generated Code - Part I}
\label{table:11a}
\begin{tabular}{cllrr}
\toprule
\textbf{Rank} & \textbf{Category} & \textbf{Violation Type} & \textbf{Count} & \cellcolor{gray!30}\textbf{Percentage} \\
\midrule
\multicolumn{5}{c}{\cellcolor{gray!25}GPT-3.5-turbo}  \\
\midrule
1 & $\mathcal{BO}$ & Buffer overflow on \texttt{scanf} & $84\,213$ & \cellcolor{gray!30}38.23\% \\
2 & $\mathcal{DF}$ & Dereference failure: NULL pointer & $56\,690$ & \cellcolor{gray!30}25.74\% \\
3 & $\mathcal{DF}$ & Dereference failure: invalid pointer & $20\,617$ & \cellcolor{gray!30}9.36\% \\
4 & $\mathcal{DF}$ & Dereference failure: forgotten memory & $4\,631$ & \cellcolor{gray!30}2.10\% \\
5 & $\mathcal{DF}$ & Array bounds violated: lower bound & $8\,102$ & \cellcolor{gray!30}3.68\% \\
6 & $\mathcal{DF}$ & Array bounds violated: upper bound & $8\,101$ & \cellcolor{gray!30}3.68\% \\
7 & $\mathcal{AO}$ & Arithmetic overflow on sub & $6\,627$ & \cellcolor{gray!30}3.01\% \\
8 & $\mathcal{DF}$ & Dereference failure: array bounds violated & $6\,537$ & \cellcolor{gray!30}2.97\% \\
9 & $\mathcal{AO}$ & Arithmetic overflow on add & $5\,228$ & \cellcolor{gray!30}2.37\% \\
10 & $\mathcal{AO}$ & Arithmetic overflow on mul & $4\,285$ & \cellcolor{gray!30}1.95\% \\
\midrule
\multicolumn{5}{c}{\cellcolor{gray!25}Falcon-180B} \\
\midrule
1 & $\mathcal{BO}$ & Buffer overflow on \texttt{scanf} & $49\,175$ & \cellcolor{gray!30}34.37\% \\
2 & $\mathcal{DF}$ & Dereference failure: NULL pointer & $42\,177$ & \cellcolor{gray!30}29.48\% \\
3 & $\mathcal{DF}$ & Dereference failure: invalid pointer & $15\,732$ & \cellcolor{gray!30}11.00\% \\
4 & $\mathcal{DF}$ & Dereference failure: forgotten memory & $5\,442$ & \cellcolor{gray!30}3.80\% \\
5 & $\mathcal{DF}$ & Dereference failure: array bounds violated & $4\,545$ & \cellcolor{gray!30}3.18\% \\
6 & $\mathcal{DF}$ & Array bounds violated: upper bound & $4\,310$ & \cellcolor{gray!30}3.01\% \\
7 & $\mathcal{AO}$ & Arithmetic overflow on sub & $3\,315$ & \cellcolor{gray!30}2.32\% \\
8 & $\mathcal{DF}$ & Array bounds violated: lower bound & $3\,611$ & \cellcolor{gray!30}2.52\% \\
9 & $\mathcal{AO}$ & Arithmetic overflow on add & $2\,858$ & \cellcolor{gray!30}2.00\% \\
10 & $\mathcal{BO}$ & Buffer overflow on \texttt{fscanf} & $2\,532$ & \cellcolor{gray!30}1.77\% \\
\midrule
\multicolumn{5}{c}{\cellcolor{gray!25}Llama2-13B} \\

\midrule
1 & $\mathcal{DF}$ & Dereference failure: NULL pointer & $17\,630$ & \cellcolor{gray!30}54.45\% \\
2 & $\mathcal{DF}$ & Dereference failure: invalid pointer & $3\,089$ & \cellcolor{gray!30}9.54\% \\
3 & $\mathcal{BO}$ & Buffer overflow on \texttt{scanf} & $2\,775$ & \cellcolor{gray!30}8.57\% \\
4 & $\mathcal{DF}$ & Dereference failure: array bounds violated & $1\,611$ & \cellcolor{gray!30}4.98\% \\
5 & $\mathcal{DF}$ & Dereference failure: forgotten memory & $1\,254$ & \cellcolor{gray!30}3.87\% \\
6 & $\mathcal{AO}$ & Arithmetic overflow on add & $883$ & \cellcolor{gray!30}2.73\% \\
7 & $\mathcal{DF}$ & Array bounds violated: upper bound & $818$ & \cellcolor{gray!30}2.53\% \\
8 & $\mathcal{AO}$ & Arithmetic overflow on mul & $599$ & \cellcolor{gray!30}1.85\% \\
9 & $\mathcal{AO}$ & Arithmetic overflow on sub & $571$ & \cellcolor{gray!30}1.76\% \\
10 & $\mathcal{BO}$ & Division by zero & 462 & \cellcolor{gray!30}1.43\% \\

\midrule
\multicolumn{5}{c}{\cellcolor{gray!25}Gemma-7B} \\
\midrule
1 & $\mathcal{DF}$ & Dereference failure: NULL pointer & $59\,433$ & \cellcolor{gray!30}60.50\% \\
2 & $\mathcal{BO}$ & Buffer overflow on \texttt{scanf} & $14\,950$ & \cellcolor{gray!30}15.22\% \\
3 & $\mathcal{DF}$ & Dereference failure: invalid pointer & $3\,617$ & \cellcolor{gray!30}3.68\% \\
4 & $\mathcal{DF}$ & Dereference failure: forgotten memory & $3\,191$ & \cellcolor{gray!30}3.25\% \\
5 & $\mathcal{DF}$ & Array bounds violated: upper bound & $3\,379$ & \cellcolor{gray!30}3.44\% \\
6 & $\mathcal{DF}$ & Array bounds violated: lower bound & $2\,784$ & \cellcolor{gray!30}2.83\% \\
7 & $\mathcal{AO}$ & Arithmetic overflow on sub & $2\,040$ & \cellcolor{gray!30}2.08\% \\
8 & $\mathcal{DF}$ & Dereference failure: array bounds violated & $1\,786$ & \cellcolor{gray!30}1.82\% \\
9 & $\mathcal{AO}$ & Arithmetic overflow on floating-point ieee\_mul & $1\,152$ & \cellcolor{gray!30}1.17\% \\
10 & $\mathcal{AO}$ & Arithmetic overflow on add & $1\,302$ & \cellcolor{gray!30}1.33\% \\

\bottomrule
\end{tabular}
\end{table}

\begin{table}[htbp]
\centering
\renewcommand{\thetable}{11 (Cont.)}
\caption{Top 10 Vulnerabilities in LLM Generated Code - Part II}
\label{table:11b}
\begin{tabular}{cllrr}
\toprule
\textbf{Rank} & \textbf{Category} & \textbf{Violation Type} & \textbf{Count} & \cellcolor{gray!30}\textbf{Percentage} \\
\midrule
\multicolumn{5}{c}{\cellcolor{gray!25}CodeLlama-13B} \\
\midrule
1 & $\mathcal{DF}$ & Dereference failure: NULL pointer & $11\,546$ & \cellcolor{gray!30}44.75\% \\
2 & $\mathcal{BO}$ & Buffer overflow on \texttt{scanf} & $5\,169$ & \cellcolor{gray!30}20.03\% \\
3 & $\mathcal{DF}$ & Dereference failure: invalid pointer & $3\,481$ & \cellcolor{gray!30}13.49\% \\
4 & $\mathcal{DF}$ & Dereference failure: array bounds violated & $897$ & \cellcolor{gray!30}3.48\% \\
5 & $\mathcal{DF}$ & Dereference failure: forgotten memory & $695$ & \cellcolor{gray!30}2.69\% \\
6 & $\mathcal{DF}$ & Array bounds violated: upper bound & $683$ & \cellcolor{gray!30}2.65\% \\
7 & $\mathcal{AO}$ & Arithmetic overflow on add & $524$ & \cellcolor{gray!30}2.03\% \\
8 & $\mathcal{DF}$ & Array bounds violated: lower bound & $465$ & \cellcolor{gray!30}1.80\% \\
9 & $\mathcal{AO}$ & Arithmetic overflow on mul & $456$ & \cellcolor{gray!30}1.77\% \\
10 & $\mathcal{AO}$ & Arithmetic overflow on sub & $380$ & \cellcolor{gray!30}1.47\% \\
\midrule
\multicolumn{5}{c}{\cellcolor{gray!25}Gemini Pro 1.0} \\
\midrule
1 & $\mathcal{DF}$ & Dereference failure: NULL pointer & $65\,376$ & \cellcolor{gray!30}55.95\% \\
2 & $\mathcal{DF}$ & Dereference failure: invalid pointer & $13\,272$ & \cellcolor{gray!30}11.36\% \\
3 & $\mathcal{BO}$ & Buffer overflow on \texttt{scanf} & $12\,948$ & \cellcolor{gray!30}11.08\% \\
4 & $\mathcal{DF}$ & Dereference failure: array bounds violated & $4\,250$ & \cellcolor{gray!30}3.64\% \\
5 & $\mathcal{DF}$ & Dereference failure: forgotten memory & $3\,340$ & \cellcolor{gray!30}2.86\% \\
6 & $\mathcal{AO}$ & Arithmetic overflow on mul & $2\,466$ & \cellcolor{gray!30}2.11\% \\
7 & $\mathcal{DF}$ & Array bounds violated: upper bound & $2\,285$ & \cellcolor{gray!30}1.96\% \\
8 & $\mathcal{DF}$ & Array bounds violated: lower bound & $1\,952$ & \cellcolor{gray!30}1.67\% \\
9 & $\mathcal{AO}$ & Arithmetic overflow on sub & $1\,899$ & \cellcolor{gray!30}1.63\% \\
10 & $\mathcal{AO}$ & Arithmetic overflow on add & $1\,895$ & \cellcolor{gray!30}1.62\% \\
\midrule
\multicolumn{5}{c}{\cellcolor{gray!25}Mistral-7B} \\
\midrule
1 & $\mathcal{DF}$ & Dereference failure: NULL pointer & $6\,294$ & \cellcolor{gray!30}33.17\% \\
2 & $\mathcal{BO}$ & Buffer overflow on \texttt{scanf} & $5\,125$ & \cellcolor{gray!30}27.01\% \\
3 & $\mathcal{DF}$ & Dereference failure: invalid pointer & $2\,460$ & \cellcolor{gray!30}12.97\% \\
4 & $\mathcal{DF}$ & Dereference failure: array bounds violated & $738$ & \cellcolor{gray!30}3.89\% \\
5 & $\mathcal{DF}$ & Array bounds violated: lower bound & $622$ & \cellcolor{gray!30}3.28\% \\
6 & $\mathcal{AO}$ & Arithmetic overflow on sub & $473$ & \cellcolor{gray!30}2.49\% \\
7 & $\mathcal{DF}$ & Array bounds violated: upper bound & $453$ & \cellcolor{gray!30}2.39\% \\
8 & $\mathcal{DF}$ & Dereference failure: forgotten memory & $414$ & \cellcolor{gray!30}2.18\% \\
9 & $\mathcal{AO}$ & Arithmetic overflow on add & $400$ & \cellcolor{gray!30}2.11\% \\
10 & $\mathcal{BO}$ & Buffer overflow on \texttt{sscanf} & $388$ & \cellcolor{gray!30}2.04\% \\
\midrule
\multicolumn{5}{c}{\cellcolor{gray!25}GPT-4o-mini} \\
\midrule
1 & $\mathcal{BO}$ & Buffer overflow on \texttt{scanf} & $33\,307$ & \cellcolor{gray!30}42.60\% \\
2 & $\mathcal{DF}$ & Dereference failure: NULL pointer & $17\,539$ & \cellcolor{gray!30}22.43\% \\
3 & $\mathcal{DF}$ & Dereference failure: invalid pointer & $7\,055$ & \cellcolor{gray!30}9.02\% \\
4 & $\mathcal{AO}$ & Arithmetic overflow on sub & $2\,479$ & \cellcolor{gray!30}3.17\% \\
5 & $\mathcal{DF}$ & Array bounds violated: upper bound & $2\,277$ & \cellcolor{gray!30}2.91\% \\
6 & $\mathcal{AO}$ & Arithmetic overflow on add & $2\,114$ & \cellcolor{gray!30}2.70\% \\
7 & $\mathcal{DF}$ & Dereference failure: array bounds violated & $1\,956$ & \cellcolor{gray!30}2.50\% \\
8 & $\mathcal{AO}$ & Arithmetic overflow on floating-point ieee\_mul & $1\,857$ & \cellcolor{gray!30}2.38\% \\
9 & $\mathcal{AO}$ & Arithmetic overflow on mul & $1\,536$ & \cellcolor{gray!30}1.96\% \\
10 & $\mathcal{DF}$ & Array bounds violated: lower bound & $1\,398$ & \cellcolor{gray!30}1.79\% \\
\bottomrule
\end{tabular}
\end{table}

\begin{table}[htbp]
\centering
\renewcommand{\thetable}{11 (Cont.)}
\caption{Top 10 Vulnerabilities in LLM Generated Code - Part III}
\label{table:11c}
\begin{tabular}{cllrr}
\toprule
\textbf{Rank} & \textbf{Category} & \textbf{Violation Type} & \textbf{Count} & \cellcolor{gray!30}\textbf{Percentage} \\
\midrule
\multicolumn{5}{c}{\cellcolor{gray!25}Falcon2-11B}  \\
\midrule
1 & $\mathcal{DF}$ & Dereference failure: NULL pointer & $12\,863$ & \cellcolor{gray!30}40.71\% \\
2 & $\mathcal{BO}$ & Buffer overflow on \texttt{scanf} & $6\,593$ & \cellcolor{gray!30}20.87\% \\
3 & $\mathcal{DF}$ & Dereference failure: invalid pointer & $4\,515$ & \cellcolor{gray!30}14.29\% \\
4 & $\mathcal{DF}$ & Dereference failure: array bounds violated & $1\,266$ & \cellcolor{gray!30}4.01\% \\
5 & $\mathcal{DF}$ & Dereference failure: forgotten memory & $1\,106$ & \cellcolor{gray!30}3.50\% \\
6 & $\mathcal{DF}$ & Array bounds violated: upper bound & $1\,074$ & \cellcolor{gray!30}3.40\% \\
7 & $\mathcal{AO}$ & Arithmetic overflow on add & $762$ & \cellcolor{gray!30}2.41\% \\
8 & $\mathcal{DF}$ & Array bounds violated: lower bound & $613$ & \cellcolor{gray!30}1.94\% \\
9 & $\mathcal{AO}$ & Arithmetic overflow on sub & $561$ & \cellcolor{gray!30}1.78\% \\
10 & $\mathcal{AO}$ & Arithmetic overflow on mul & $459$ & \cellcolor{gray!30}1.45\% \\
\bottomrule
\end{tabular}
\end{table}

Dereference failures, particularly NULL pointer dereferences, are among the most prevalent vulnerabilities across all LLMs. This is due in part to the varied and often unsafe examples of pointer usage in training datasets, combined with the inherent complexity of dynamic memory management in C. LLMs rely on pattern recognition rather than deep understanding, which leads them to mishandle pointers and fail to replicate the nuanced behavior of real-world applications. This results in frequent dereference issues and flawed pointer handling, highlighting significant risks when deploying LLM-generated code in critical systems where security and reliability are paramount.

The severity and frequency of these vulnerabilities vary significantly among models. For instance, Gemma-7B exhibits a notably high rate of NULL pointer dereference failures at 60.50\%, indicating substantial weaknesses in memory management. Arithmetic overflows also consistently appear across all models in the top 10 list, and differ based on specific operations (addition, subtraction, multiplication), underscoring varied arithmetic handling. Notably, Llama2-13B stands out with less than 10\% of \texttt{scanf} violations, with Gemini Pro 1.0 close behind at approximately 11\%; however, both models, like Gemma-7B, show high rates of NULL pointer dereference failures.

The consistent occurrence of certain errors across different models underscores the need for comprehensive testing and validation frameworks to address these recurring issues before deployment. While all models share similar vulnerabilities, significant differences in the frequency and types of other vulnerabilities—such as arithmetic overflows—suggest that model-specific optimizations and enhancements are necessary.
To mitigate these risks, developing enhanced training methodologies focused on robust memory handling is crucial. Implementing advanced code analysis tools and frameworks is also essential to detect and rectify vulnerabilities before deployment for real-world applications.
\subsection{LLM Ranking: Which model is the most secure coder}

To compare which model is performing the ``worst'' or the ``best'' when it comes to secure coding---and to do this as fairly as possible---we will investigate several metrics, such as the ratio of verification results, average property violation per file, and average property violation per line of code.

\addtocounter{table}{-2}  
\renewcommand{\thetable}{\arabic{table}} 

\begin{table}[b]
\centering
\caption{Verification Results Summary, Sorted by Average Property Violation per Line.}
\label{tab:verification_results}
\begin{tabular}{@{}lccccccc@{}}
\toprule
Category & \cellcolor{yellow!20}\begin{tabular}[c]{@{}c@{}}Avg Prop.\\ Viol.\\ per Line\end{tabular} & \cellcolor{green!20}Rank & \cellcolor{green!20} \begin{tabular}[c]{@{}c@{}}$\mathcal{VS}$\end{tabular} & \cellcolor{red!30}Rank & \cellcolor{red!30} \begin{tabular}[c]{@{}c@{}}$\mathcal{VF}$\end{tabular} & \begin{tabular}[c]{@{}c@{}}$\mathcal{VU}$\\ (Timeout)\end{tabular} & \cellcolor{gray!30} \begin{tabular}[c]{@{}c@{}}Avg Prop.\\ Viol.\\ per File\end{tabular} \\ \midrule
GPT-4o-mini & \cellcolor{yellow!20}\textbf{0.0165} & \cellcolor{green!20}3 & \cellcolor{green!20}4.23\% & \cellcolor{red!30}2 & \cellcolor{red!30}57.14\% & 36.77\% & \cellcolor{gray!30}3.40 \\
Llama2-13B & \cellcolor{yellow!20}0.0234 & \cellcolor{green!20}2 & \cellcolor{green!20}12.36\% & \cellcolor{red!30}\textbf{1} & \cellcolor{red!30}\textbf{51.30\%} & 31.78\% & \cellcolor{gray!30}3.62 \\
Mistral-7B & \cellcolor{yellow!20}0.0254 & \cellcolor{green!20}7 & \cellcolor{green!20}8.36\% & \cellcolor{red!30}4 & \cellcolor{red!30}62.08\% & 25.88\% & \cellcolor{gray!30}\textbf{3.07} \\
CodeLlama-13B & \cellcolor{yellow!20}0.0260 & \cellcolor{green!20}\textbf{1} & \cellcolor{green!20}\textbf{15.48\%} & \cellcolor{red!30}3 & \cellcolor{red!30}52.71\% & 29.52\% & \cellcolor{gray!30}4.13 \\
Falcon-180B & \cellcolor{yellow!20}0.0291 & \cellcolor{green!20}8 & \cellcolor{green!20}6.48\% & \cellcolor{red!30}5 & \cellcolor{red!30}62.07\% & 28.67\% & \cellcolor{gray!30}3.38 \\
GPT-3.5-turbo & \cellcolor{yellow!20}0.0295 & \cellcolor{green!20}6 & \cellcolor{green!20}7.29\% & \cellcolor{red!30}7 & \cellcolor{red!30}65.07\% & 26.09\% & \cellcolor{gray!30}4.42 \\
Gemini Pro 1.0 & \cellcolor{yellow!20}0.0305 & \cellcolor{green!20}5 & \cellcolor{green!20}9.49\% & \cellcolor{red!30}6 & \cellcolor{red!30}63.91\% & 24.13\% & \cellcolor{gray!30}4.70 \\
Gemma-7B & \cellcolor{yellow!20}0.0437 & \cellcolor{green!20}4 & \cellcolor{green!20}11.62\% & \cellcolor{red!30}8 & \cellcolor{red!30}67.01\% & 16.30\% & \cellcolor{gray!30}4.20 \\
\bottomrule
\end{tabular}
\smallskip 
\center
\textbf{Legend:} \\
$\boldsymbol{\mathcal{VS}}$: Verification Success;
$\boldsymbol{\mathcal{VF}}$: Verification Failed;
$\boldsymbol{\mathcal{VU}}$: Verification Unknown (Timeout).\\ Best performance in a category is highlighted with bold and/or Rank.
\end{table}

\textbf{The results indicate that there is no clear winner.} Mistral-7B, despite having the fewest property violations per file, writes shorter code, reducing its likelihood of coding errors. However, this model also performs poorly in the $\mathcal{VS}$ metric, with only 8.36\% of its samples categorized as being free of vulnerabilities. CodeLlama-13B achieved the highest $\mathcal{VS}$ rate, followed by Llama2-13B, and their $\mathcal{VF}$ ratio ranking is third and second respectively, which is a good result for the Llama family. Still, it is best to remember that nearly half of their samples had vulnerabilities. Moreover, their $\mathcal{VU}$ is fairly high at $30\%$ and $32\%$, which means that with further verification, there is still a chance that other models will take the lead.

GPT-4o-mini outperforms GPT-3.5-turbo while showing the highest $\mathcal{VU}$ percentage under current ESBMC settings, indicating its ability to produce more complex and longer outputs. It is important to note that this complexity is not reflected by the CC number as discussed earlier, which confirms the criticism towards Cyclomatic Complexity by practitioners. While GPT-4o-mini ranks third in $\mathcal{VS}$ and second in $\mathcal{VF}$, it finishes first with an average property violation per line. This might be the fairest way to compare models, as the more lines, the more chances to have vulnerabilities, while this metric doesn't punish models producing shorter codes.
While there is no definitive winner in this analysis, Gemma-7B, Gemini-Pro, and GPT-3.5-turbo---with the current verification settings--- have the highest $\mathcal{VF}$ ratios and highest average property violation both per line and file which indicates that these models are performing worse in our test.

It is important to underline that it might be tempting to speculate on a winner, having such a high verification failed ratio is unacceptable from an SE perspective for any model. All models surpassed the $\mathcal{VF}$ threshold of 50\%, indicating that nearly half or more of the generated programs are vulnerable.
The conclusions of this analysis must be clear: \textbf{Using code generated by the state-of-the-art Large Language Models, without any additional framework for validation and vulnerability analysis, carries severe risks.} While LLMs can be useful for automating simple tasks and scripting, directly including \textbf{such codes in production software without oversight from experienced software engineers is irresponsible and should be avoided.}

\section{Limitations and Future Research}
\label{sec:futureandlimit}

\subsection{Future Research Directions}

The dataset, consisting of $331\,000$ C program files and their corresponding vulnerability classifications, is available on GitHub\footnote{\url{https://github.com/FormAI-Dataset}}. The dataset is well-suited for machine learning applications and fine-tuning LLMs due to its large size. Moreover, the diverse structure of the C programs generated by various LLMs in the FormAI-v2 dataset makes it ideal for an unexpected use case: fuzzing different applications. We discovered and reported seventeen bugs in the ESBMC application, including issues related to unsigned overflow checks, SMT solver problems, conversion errors in the GOTO converter, and flaws in implementing the \textit{k}-induction proof rule. Furthermore, we identified bugs in the CBMC~\cite{kroening2014cbmc} tool while using the FormAI-v2 dataset and promptly communicated these findings to the respective developers. After validating the reported issues, the ESBMC developers have already resolved thirteen.

Our results give rise to several interesting research directions:
\begin{itemize}
    \item It would be important to investigate why programs under ``Verification Successful'' are void of vulnerabilities. Is it because of better coding practices or simply because, for example, they don't take user input, thereby avoiding buffer overflows? 
    \item What is the right path towards LLMs producing secure code: Re-training models on better data, fine-tuning, or using current models in various few-shot frameworks with better prompting?
    \item Since several codes contain multiple vulnerabilities, this dataset is ideal for bench-marking and testing various vulnerability detection tools.
    \item As our motivation section showcased, GPT-4o-mini did not excel at avoiding and fixing the vulnerability in the example. How do different LLMs compare in understanding, correctly fixing, and detecting coding errors?
    \item We aim further to grow the FormAI dataset, including more state-of-the-art models, and increase the number of samples for each LLM to have an overall larger dataset.
    \item How do different programming Tasks or Styles impact vulnerable coding patterns? Are there tasks that LLMs consistently mess up?
\end{itemize}

While we can partially address the last question, noting the use of insecure functions and poor input sanitation in handling user inputs, exploring this issue across various domains, such as networking or cryptography, would be beneficial.

\subsection{Limitations and Threats to Validity}


ESBMC might find slightly more vulnerabilities in a given program with a larger timeout setting. Whether the verifier can finish the process under a given timeout is up to the available computational capacity. The same parameter setting can yield a higher or lower detection rate on different architectures.
To find all errors detectable by ESBMC, unwind must be set to infinite, and ESMBC must complete the verification process. As we provided the original C programs and the instructions on how to run ESBMC, researchers who invest additional computational resources have the potential to enhance our findings.
As the ``Verification Unknown'' category still contains samples for every model, the current results are strongly bound to the percentage of vulnerable files LLMs produce. 

While ESBMC is a robust tool for detecting many types of errors in C, it is not currently suited to detect design flaws, semantic errors, or performance issues. As such, more vulnerabilities might be present besides the ones detected in the code. Thus, we recommend that the training and fine-tuning applications be restricted to the vulnerabilities detectable by ESBMC on this dataset.

All programs shorter than 50 lines were removed from the dataset. However, researchers interested in smaller programs can still find all programs under 50 lines, generated by GPT-3.5-turbo, in the original FormAI-v1 dataset.

\section{Conclusions}
\label{sec:conclusion}

This research analyzed nine state-of-the-art Large Language Models to assess their likelihood of introducing vulnerabilities during neutral prompt-based code generation, and to compare their performance. The models included in our analysis were Mistral-7B, Falcon-180B, Falcon2-11B GPT-4o-mini, Llama2-13B, CodeLlama-13B, Gemma-7B, GPT-3.5-turbo, and Gemini-Pro. We employed a zero-shot prompting method to encompass numerous programming scenarios for C code generation. These programs constitute the FormAI-v2 dataset, containing $331\,000$ independent compilable C programs.

We used the Efficient SMT-based Bounded Model Checker (ESBMC), a state-of-the-art formal verification tool, to identify vulnerabilities. Each program was given a verification period of 500 seconds with the unwinding parameter set to infinite, uncovering a total of $765\,366$ vulnerabilities. Overall $62.07\%$ of the codes were vulnerable. Detailed labeling of each sample---including filename, type of vulnerability, function name, error type, and source code---is documented in a .json file, as detailed in Appendix Fig.~\ref{fig:json}, to facilitate the dataset's use in machine learning applications.

Additionally, the FormAI-v2 dataset proved instrumental for fuzzing various applications and identifying multiple bugs in ESBMC and CBMC. These findings provide clear answers to our research questions:

\begin{tcolorbox}[colback=gray!10]

\begin{itemize}
    \item {\textbf{RQ1}:} How does the security of LLM-generated code differ across various models?
    \item {\textbf{Answer}:} CodeLlama-13B, Llama-13B, and GPT-4o-mini perform slightly better, but all examined models notoriously introduce vulnerabilities into the C code they generate at unacceptable rates. Our research revealed that all examined models introduced vulnerabilities in at least 50\% of the generated code. 
    
\end{itemize}

\end{tcolorbox}

\begin{tcolorbox}[colback=gray!10]
\begin{itemize}
\item {\textbf{RQ2}:} What are the most typical vulnerabilities introduced by different LLMs during code generation (focusing on C)?
\item {\textbf{Answer}:} 
Dereference failures and buffer overflow issues are the most prevalent vulnerabilities across all models, ranking arithmetic overflow as the third most common type. No model is completely free from any of the examined vulnerabilities; the variations lie in the frequency of occurrence.
    
\end{itemize}

\end{tcolorbox}

While the literature reveals significant variations in these models' ability to solve tasks, this is not mirrored in their susceptibility to produce vulnerabilities in source code. Our findings conclusively show that despite differences among the examined models in terms of generating code, they all consistently introduce severe vulnerabilities when prompted with simple coding tasks. 
Our study indicates that despite the impressive capabilities of Large Language Models in code generation, employing their output in production requires detailed risk assessment. Relying on these models without expert oversight in a production context is inadvisable.

\section*{Acknowledgement}
We extend our sincere thanks to the anonymous reviewers for their valuable feedback, which has significantly improved the quality of this paper. This research is supported by the Technology Innovation Institute (TII), Abu Dhabi. Additionally, partial support is provided by the EPSRC grant EP/T026995/1, titled ``EnnCore: End-to-End Conceptual Guarding of Neural Architectures'' under the Security for All in an AI-enabled Society initiative. This work is also partially supported by the TKP2021-NVA Funding Scheme under Project TKP2021-NVA-29.

\section*{Data Availability Statements}

This study generated and examined a total of $331\,000$ C samples. The findings and all the generated C samples are available for access and download from the project's website at \url{https://github.com/FormAI-Dataset}.

\section*{Conflicts of interest}
The authors have no competing interests to declare that are relevant to the content of this
article.

\bibliography{sn-bibliography}

\section*{Appendix}
\begin{appendices}

\begin{figure}
    \centering
    \includegraphics[width=1\linewidth]{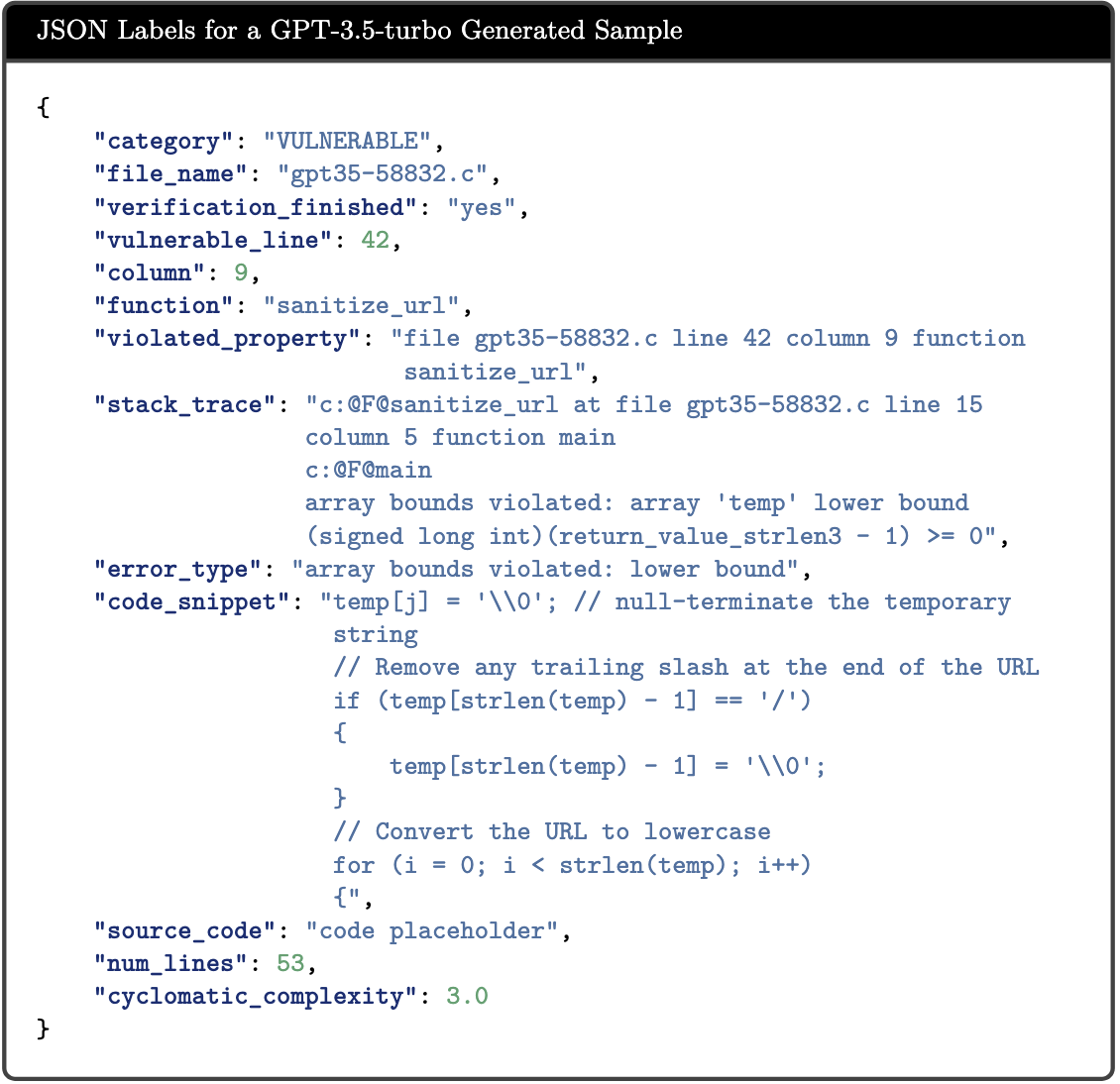}
    \caption{Example JSON Labels for a GPT-3.5-turbo Generated Sample: FormAI-v2 dataset}
    \label{fig:json}
\end{figure}

\end{appendices}

\end{document}